\documentclass[preprints,article,accept,moreauthors,pdftex,10pt,a4paper]{mdpi}
 
\usepackage{amsmath,amsfonts,amssymb}
\usepackage{graphicx}

\firstpage{1} 
\pubvolume{xx}
\issuenum{xx}
\articlenumber{xx}
\pubyear{2019}
\copyrightyear{2019}
\history{February 28, 2019}

\Title{Two-dimensional electronic transport in 
{
rubrene: the impact of inter-chain coupling}}

\Author{Ahmed Missaoui $^{1,3,\ddagger}$\orcidA{}, Jouda Jemaa Khabthani $^{2,\ddagger}$\orcidB{}, Guy {Trambly de Laissardi\`ere} $^{3,\ddagger}$*\orcidC{} and Didier Mayou $^{4,\ddagger}$\orcidD{}}

\AuthorNames{Ahmed Missaoui, Jouda Jemaa Khabthani, Guy Trambly de Laissardi\`ere and Didier Mayou}

\address{%
$^{1}$ \quad Laboratoire de Spectroscopie Atomique Mol\'eculaire et Applications, 
D\'epartement de Physique, Facult\'e des Sciences de Tunis, Universit\'e de Tunis El Manar,  Campus universitaire 1060 Tunis, Tunisia; ahmed.missaoui@fst.utm.tn\\
$^{2}$ \quad Laboratoire de la Physique de la Mati\`ere Condens\'ee, 
D\'epartement de Physique, Facult\'e des Sciences de Tunis, Universit\'e de Tunis El Manar,  Campus universitaire 1060 Tunis, Tunisia; jouda.khabthani@fst.utm.tn\\
$^{3}$ \quad Laboratoire de Physique th\'eorique et Mod\'elisation, CNRS and Universit\'e de Cergy-Pontoise, 
95302 Cergy-Pontoise, France; guy.trambly@u-cergy.fr\\
$^{4}$ \quad CNRS -  Universit\'e Grenoble Alpes, Inst NEEL, F-38042 Grenoble, France; didier.mayou@neel.cnrs.fr \\
}

\corres{Correspondence: guy.trambly@u-cergy.fr; Tel.: +33-1-3425-7523}

\secondnote{These authors contributed equally to this work.}

\begin{document}
\abstract{abstract}

Abstract: Organic semi-conductors have unique electronic properties and are important  systems both at the fundamental level and also for their applications in electronic devices. In this article we focus on the particular  case of rubrene which has one of the best electronic transport properties for application purposes. We show that this system can be well simulated by simple tight-binding systems representing one-dimensional (1D) chains that are weakly coupled to their neighboring chains in the same plane. This makes in principle this rubrene  system somehow intermediate between 1D and isotropic 2D models. We analyse in detail the dc-transport and terahertz conductivity in the 1D and in the anisotropic 2D models. The transient localisation scenario allows us to reproduce satisfactorily some basics results such as mobility anisotropy and orders of magnitude as well as ac-conductivity in the terahertz range. This model shows in particular that even a weak inter-chain coupling is able to improve notably the propagation along the chains. This suggest also that a strong inter-chain coupling is important to get  organic semi-conductors with the best possible transport properties for applicative purposes.

\tableofcontents
\section{Introduction}

 In 1977, Alan J. Heeger, Alan G. MacDiarmid and Hideki Shirakawa \cite{6C.K} showed that doped polymers can conduct electricity and possess relatively high room temperature conductivities of the order of a few hundred Ohm$^{-1}$cm$^{-1}$. 
They have been awarded the Nobel prize in chemistry in 2000 for their discovery.
Since then several devices have been built from these materials,  taking advantage of their conduction properties \cite{6F.J.M,Chengliang12,Jo15,Le17,Ly18},  and present a great interest from an industrial point of view. Commercial products like  organic solar cells  (OPVCs), organic field effect transistors (OFETs) \cite{6C.D} and organic light emitting diodes  (OLED) are now very much used in new technologies for mobile phones, touch screens etc.
The advantages of these materials compared to others are
their low cost, their mechanical properties and in particular their flexibility and the possibility to adapt their electronic properties by modifying the molecular structures. 
 
The charge transport mechanisms in organic semi-conductors are complex. They depend on fundamental interactions like electron-phonon interaction and on the possible existence of complex quasiparticles like polarons. Disorder related to thermal motion or to imperfections in the structure play also a strong role. 
Developing theories and understanding  of the  experimental data on transport and optical properties is still a challenge.

In this paper we present a description of quantum diffusion within a scenario that is known as transient localisation scenario \cite{6Ciuchi2011,6Fartini2016}. This scenario  gives so far one of the best description \cite{Fratini17} of transport in a system like rubrene  C$_{42}$H$_{28}$ which is known to possess one of the highest mobility at room temperature among organic semi-conductors. 
{
In the transient localization scenario the polaronic effect is considered as negligible  and the charge carriers move in a  system which is described by a tight-binding Hamiltonian. The hopping matrix element of this Hamiltonian are time dependent because of the motion of the molecules. This transient localization scenario is not expected to apply to systems with much lower mobilities for which polaronic effects can be important.}
{
Our goal is to show for rubrene and, within the transient localisation scenario \cite{6Fartini2016,Fratini17}, how the structure and the interchain coupling can affect the electronic conduction.} Indeed rubrene is composed of one dimensional chains that are weakly coupled and we consider 1D and 2D models of rubrene crystals that are the two type of models used in the literature \cite{6Troisi2007,Gargiulo11,6troisi2011}. In the case of the 2D model we also consider two ways of describing disorder due to thermal motion.
{
In practice, we calculate numerically the optical conductivity of the rubrene by using a efficient real space method that takes into account static disorder and thermal motion disorder. 
We compare a 1D model --with intra-chain coupling only-- and a 2D model --with intra- and inter-chain couplings-- and we show how inter-chain couplings increase the conductivity along the rubrene chain. 
We also show that, qualitatively, a correlated disorder and an uncorrelated disorder give similar results. 
In 2D model, the calculated ratio between mobility parallel to the chains and mobility perpendicular to the chains is close to the experimental value. 
}

{
In section \ref{Sec_mat} the basic structural and electronic properties of organic semi-conductor are presented, with a special focus on rubrene. 
The theoretical model Hamiltonian used is presented section \ref{Sec_theoMod}. 
In section \ref{sec_transport}, the optical conductivity, calculated in the framework of Transient localisation model for charge transport, is presented for 1D and 2D models, with correlated or uncorrelated disorder model. 
}

\section{
{
The organic semi-conductors}
}

\label{Sec_mat}

\subsection{General aspects}

\begin{figure}[h]
\centering

\includegraphics[width=0.8\linewidth]{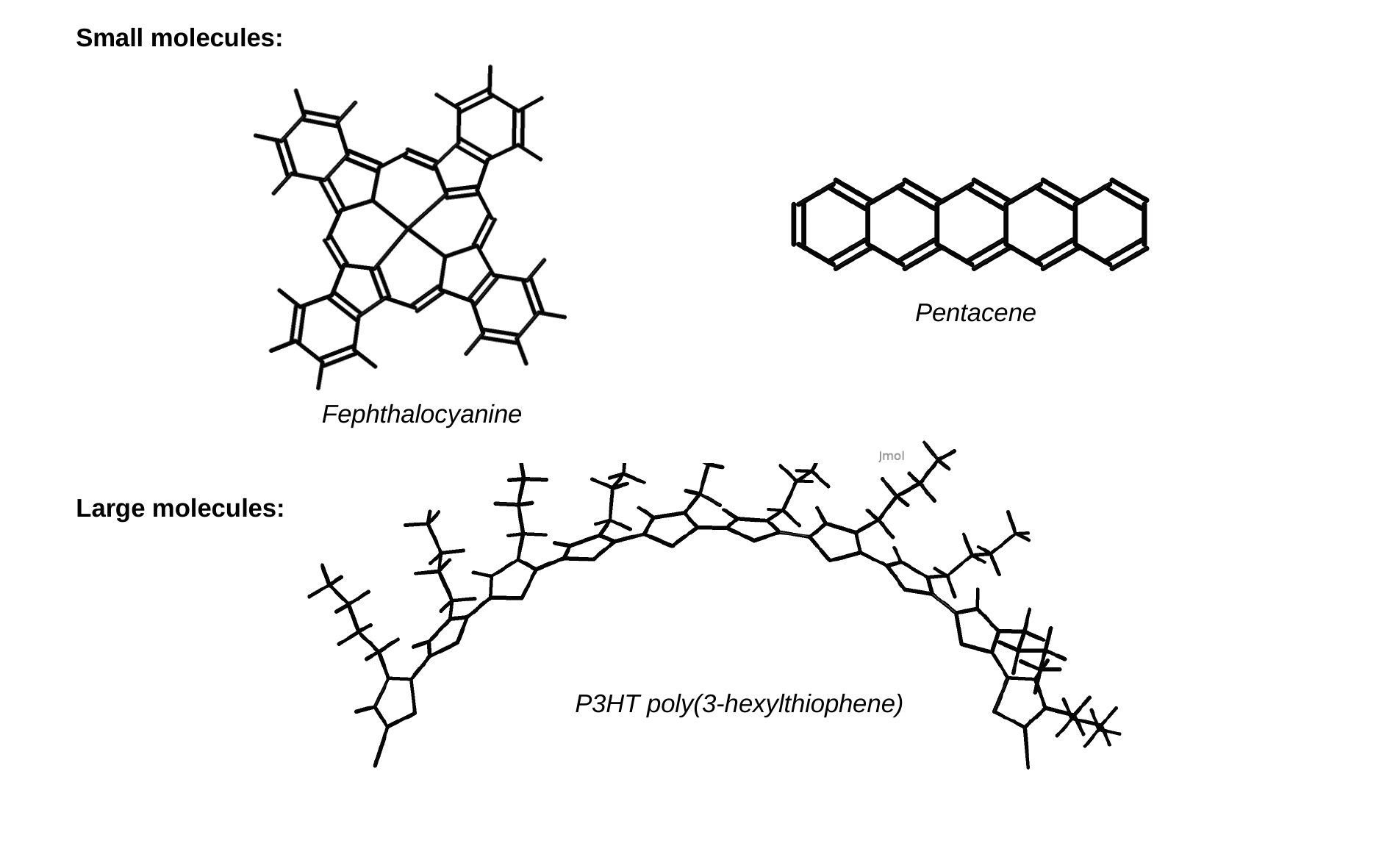}

\caption{Examples of organic semi-conductors: 
{
small and large molecules. 
}}
\label{6sco}
\end{figure}

{
Organic semi-conductors can be  classified into two groups based on their weight: large and small molecules (figure \ref{6sco}).
The basic chemical element for organic semi-conductors is of course the carbon atom.  
}
The valence electrons determine the electrical and optical properties of materials. 
They participate to the formation of chemical bond with other atoms. 
Both have in common a conjugated p-electron system formed by the p$_z$
orbitals of sp$^2$-hybridized C atoms in the molecules. The backbone of the molecules is constituted by $\sigma$ bonds, 
that are in the plane of the molecule, and the p$_{z}$ orbitals are perpendicular to the  plane of the molecule. 
The electrons in $\sigma$ orbitals are strongly localized between two neighboring atoms and cannot propagate beyond, 
while the electrons in p$_{z}$ orbitals (also called $\pi$ orbitals) are delocalized over the whole molecule. Therefore  $\pi$ orbitals can lead to a conducting band of electronic states. 
Organic semi-conductors possesses an electrical conductivity that is between that of inorganic semi-conductors and that of insulators.

\begin{figure}[h]
\centering
\includegraphics[width=0.25\linewidth]{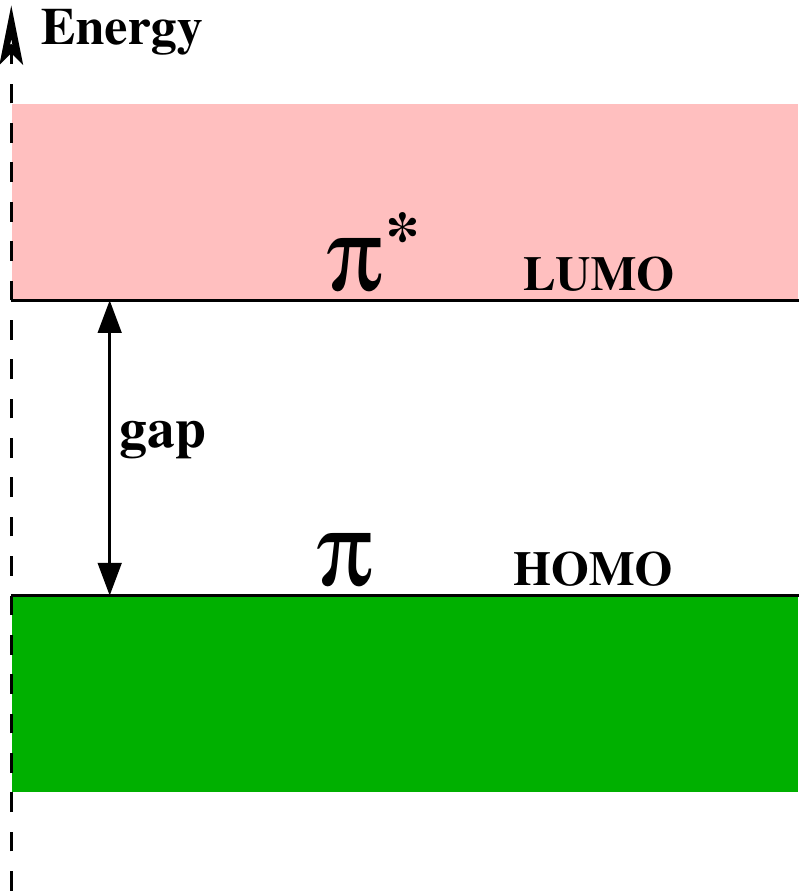}
\caption{Sketch of the energies of LUMO and HOMO orbitals.}
\label{6lumo}
\end{figure}

For electronic properties of organic semi-conductors the most pertinent states are the Lowest Unoccupied Molecular Orbital (LUMO) and the Highest Occupied Molecular Orbital (HOMO)
(figure \ref{6lumo}). As compared to energies of the HOMO and LUMO orbitals the  $\sigma$ orbitals have either much lower energy (occupied states) or much higher energy (unoccupied states). Therefore the HOMO and LUMO orbitals are linear combinations of  $\pi$ orbitals.

When organic molecules are assembled to form a crystal, the force that maintains them together is essentially the Van der Waals force. Compared to covalent bonds and interactions in inorganic semi-conductors or to bonding in metals the Van der Waals force is much smaller. That is why the properties of organic semi-conductors are very different from those of inorganic semi-conductors. Indeed the electronic properties of inorganic semi-conductors can often be explained in terms of the theory of bands. In the case of organic materials the wave-functions of the HOMO and LUMO states of every molecule can be coupled to those of the neighboring molecules. Therefore the HOMO orbitals form the valence band and the LUMO orbitals form the conduction band. Yet because of the small overlap between orbitals of different molecules (resulting from the small Van der Waals interaction) the electron band (conduction band derived from LUMO) and the hole band (valence band derived from HOMO) have very small width smaller than an eV. This small width makes the organic crystal particularly sensitive to the effect of disorder and it becomes difficult to develop a good theory of conduction and optical properties \cite{6Brocks2004}. The presence of narrow bands deeply influences the nature of charge transport in organic material.
In organic molecules used to produce semi-conducting materials the difference in energy between the LUMO and HOMO orbitals is usually of the order 1 eV to 4 eV \cite{6pope}. It corresponds to the size of the electronic gap which is therefore relatively high. This reduces the density of  thermally activated charge carriers in pure organic crystals, as compared to inorganic semi-conductors like silicon for example.

\begin{figure}[h]
\centering
\includegraphics[width=0.85\linewidth]{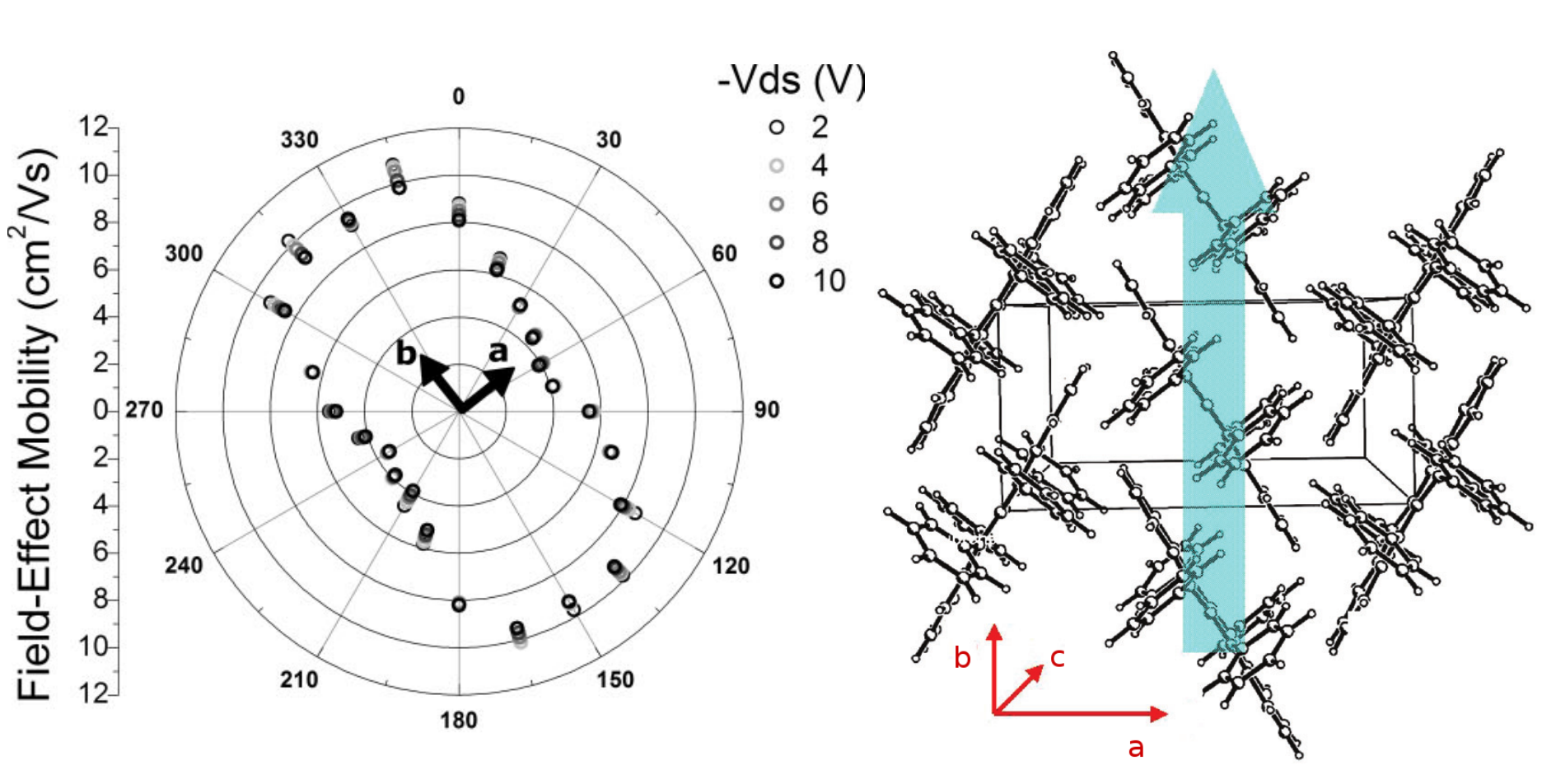}

\caption{Left panel: Polar graphics showing the mobility in a monocrystal of rubrene along different directions in the $(a,b)$ plane. The different symbols indicate the different measures \cite{6Reese2007}.
Right panel: Structure orthorhombic rubrene.
The arrow indicates the direction of the greatest mobility \cite{Sundar04}.}
\label{6mobilit}
\end{figure}

The nature of charge transport in organic semi-conductors is still an open question. A challenge for theorists is to develop a model that is adequate to describe the temperature dependence of the mobility. Experiments performed on naphtalene and anthracene \cite{6Karl2003,6Karl2001} show that the mobility decreases rapidly when temperature increases and shows also an anisotropy along the different crystallographic directions (figure \ref{6mobilit} ). This decrease of the mobility comes essentially from the effect of phonons. Many models exist to explain this behavior but they usually encounter difficulties either in the low or high temperature range.

One can cite for example the model of band transport which is based on the fact that the electronic states are delocalized like plane waves and are scattered by phonons leading to a mean-free path $L_{e}$, $L_{e}\gg a$ (with $a$ the intermolecular  distance), and a large mobility $\mu\gg 1$ cm$^{2}/$Vs. This model is not valid at high temperature (greater than room temperature) where scattering by phonons becomes important. There are also models based on hopping \cite{6POLLAK1991}. Here the idea is that states are essentially localized and transport is allowed by hopping produced by interaction with phonons. In that case the mean-free path is short $L_{e}\ll a$ and $\mu\ll 1$ cm$^{2}/$Vs. This model cannot be valid at low temperature where mobilities are high. The understanding of the transport mechanism must in particular take into account several important aspects including the effect of molecular vibrations on the hopping integrals.  

\subsection{The rubrene}

\begin{figure}[h]
  \centering
  \includegraphics[width=5cm]{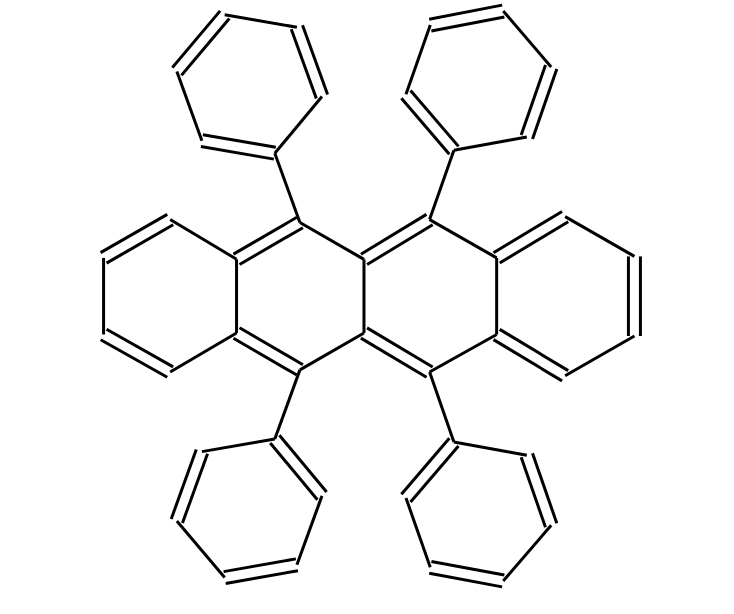}~~~~~~~\includegraphics[width=4cm]{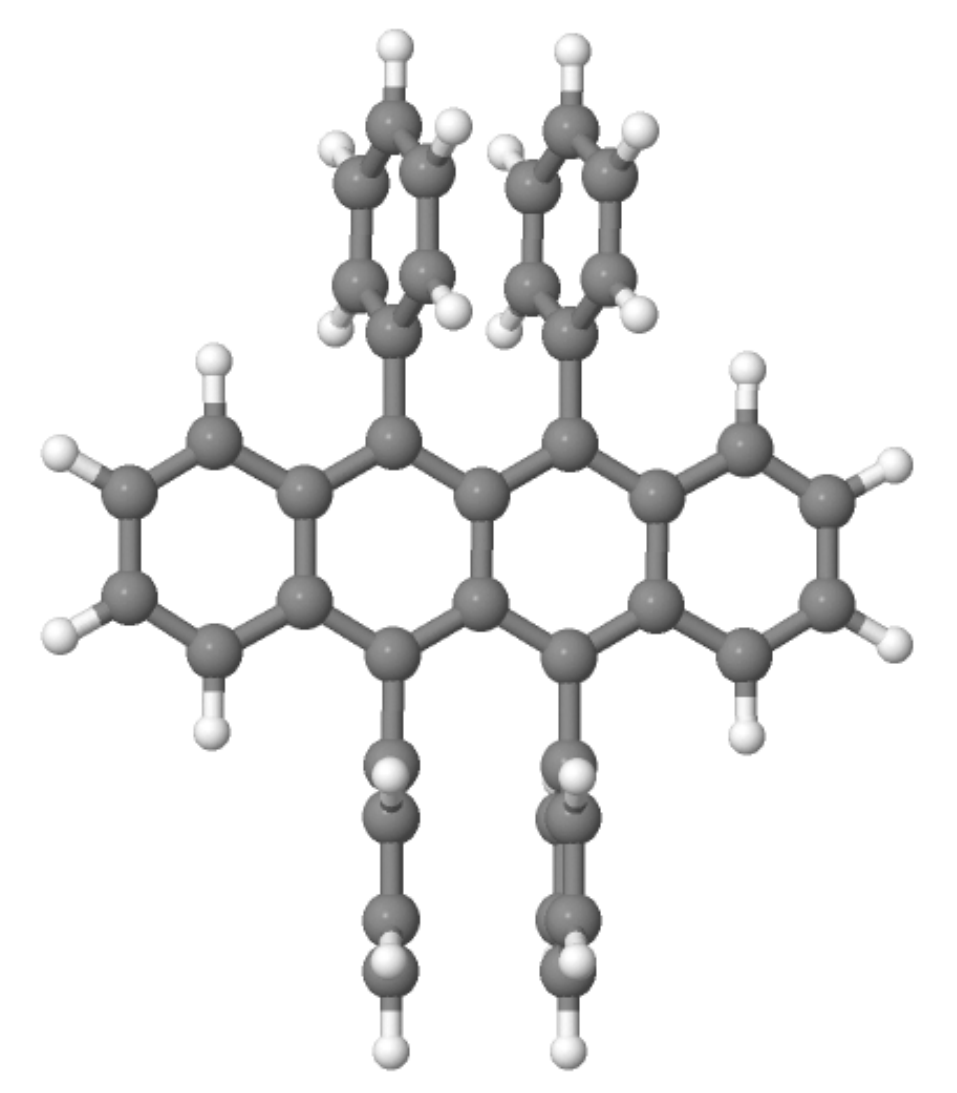}
   \caption{Left panel: Structural formula of the rubrene molecule. Right panel: 3D representation of the rubrene molecule. Black disks are carbon atoms and white disks are hydrogen atoms.}
   \label{6crsito}
\end{figure}

As a prototypical organic molecule, rubrene  has attracted considerable attention due to its high photoluminescent yield of nearly $100\%$ \cite{Uchida99,6Gosuke1999}.
Its formula is $\rm C_{42}H_{28}$ 
(figure \ref{6crsito}), 
and its high charge carrier hole mobility of $3.74$\,cm$^2$Vs$^{-1}$ can be attributed to the strong $\pi$-$\pi$ overlap between the adjacent rubrene molecules \cite{Sim18}. 
It has been used in devices such as chemical sensors \cite{Botzung-Appert04}, as well as OLEDs \cite{Wang10,choi14} and OFETS  \cite{Yi12}.
The rubrene molecule (figure \ref{6crsito}) belongs to the group of polycyclic aromatics hydrocarbons (PAH). It is formed from a tetracene molecule (4 benzenic cycles in a linear arrangement) and 4 lateral phenyls. It has been shown \cite{6silva2005} that in the neutral molecule the energies and shapes of the LUMO and HOMO orbitals are very close to those of tetracene. This similarity shows that the molecular orbitals of the lateral phenyls do not participate to the HOMO and LUMO states.

The HOMO and LUMO levels of rubrene and tetracene are close and their respective values are   $-4.69$ and $-2.09$\,eV for rubrene and  $-4.87$ and $-2.09$\,eV for tetracene \cite{6silva2005}. Experimentally the ionisation energy has been measured to  $4.9$\,eV \cite{6Yasuo2012,6Yasuo2008} which corresponds to the calculated value of the  HOMO level.

The crystallisation of rubrene can be obtained by several methods with different molecular arrangements \cite{6Jurchescu2006,6huang}. The crystalline form has a key influence on the mobility of charges and on the global performance in all organic electronic devices. The rubrene crystals in the orthorhombic phase are the easiest to obtain, even with size up to a few centimeters, 
and they have the best transport properties, compared to other
polymorphes \cite{6Podzorov,6Takeshi}. Therefore in the following we only consider this orthorhombic phase.

\begin{table}[h]
\begin{center}
\begin{tabular}{cc}
        \hline
        Parameters & Values \\ 
        \hline
        a [nm]  &1.4433 \\ 
         b [nm] & 0.7193\\ 
          c [nm]&2.686\\ 
          Volume [nm$^3$]& 2.7885\\
        \hline
\end{tabular}
\caption{Parameters of the elementary cell of orthorhombic rubrene. $T=293$\,K \cite{6Jurchescu2006}.}\label{6tab}
\end{center}
\end{table}

In table \ref{6tab}, we present the parameters of the elementary cell of the orthorhombic crystal of rubrene at room temperature ($T=293$\,K). This cell is represented in 
{
figure  \ref{6mobilit} \cite{6Jurchescu2006,Sundar04}.
}
One notices that along the $c$ axis  the molecules are simply stacked layer by layer. The molecules of successive layers interact only by their phenyl groups which leads to a strong anisotropy of intermolecular electron hopping. For this reason these electrons essentially propagate in planes perpendicular to the $c$ axis. 

As already mentioned rubrene possesses remarkable transport properties. The mobility of charge carriers in orthorhombic monocrystals can be as high as 40 cm$^{2}$/Vs comparable to that in amorphous silicon  \cite{6Podzorov,6Yamagishi2007}. As for most of organic crystals,  electronic transport in rubrene is highly anisotropic. High values of  mobility are obtained only along the $b$ axis of the lattice. This is shown in the polar graphic representing the variation of the mobility in the $(a,b)$ plane of a monocrystal of rubrene in figure \ref{6mobilit} \cite{6Podzorov,6Reese2007}. The direction with the best mobility corresponds to the direction with $\pi$-$\pi$ stacking of the rubrene molecules (the aromatic cycles are face to face). An increase of pressure increases this intermolecular coupling leading to an increase of mobility \cite{6Zhenlin2005}. As most of organic semi-conductors, rubrene is a material of p-type. The values of high mobilities are only for holes and the mobilities of electrons are smaller by several orders of magnitude \cite{6Satria2010}.

\section{Theoretical model}

\label{Sec_theoMod}

\subsection{Tight-binding Hamiltonian for rubrene}

A fully quantum description of charges transport in rubrene is obviously difficult and approximations are necessary. In this work we consider a model that treats classically the position of the molecules of rubrene and quantum mechanically the electrons \cite{6Troisi2006} with the help of a tight-binding Hamiltonian
\cite{6coropceanu2007,6Eckhardt},
\begin{equation}\label{6hamil}
\hat{H} = \sum_{i}\epsilon_{i}a^{+}_{i}a_{i} + \sum_{i\neq j}J_{ij}a^{+}_{i}a_{j} \,,
\end{equation}
where $a^{+}_{i},a_{j}$ are respectively the creation and annihilation operators for an electron on molecule $i$ and $j$. $\epsilon_{i}$ is the energy  on site $i$ and $J_{ij}$ the hopping integral between orbitals $i$ and $j$. 
In rubrene crystal we consider only one orbital per molecule which represents either the HOMO orbital (for holes) or the LUMO orbital (for electrons). The hopping term $J_{ij}$ which represents the possibility for a charge to go from one molecule to a neighbor depends on the distance between the two molecules, which itself depends on time. The time dependence $J_{ij}(t)$ is imposed by the motion of the molecules which is assumed to be independent of the presence or not of a charge on it. These assumptions are justified in  \cite{6Troisi2006}:
\begin{itemize}
\item The band structure calculations show that the LUMO and HOMO bands do not mix with the bands formed from other orbitals in most of the organic semi-conductors. Therefore a single band derived from a single state per molecule is able to reproduce the states of an electron (LUMO orbital) or of a hole (HOMO orbital).
\item The phonons that are coupled to the electrons have small frequencies and can be treated classically at room temperature. 
\item A positive or negative charge on a pentacene or rubrene molecule produces minor changes in the vibrations states of the molecule  \cite{6Coropceanu2002}. This suggests that the modulation of the transfer integral  $J_{ij}$ is insensitive to the presence or absence of a charge on the molecule and depends on its distance to the neighbor molecule.
\end{itemize}
Under these conditions the tight-binding Hamiltonian (equation (\ref{6hamil})) provides a simple model to describe the electronic properties of a molecular solid
 in which
the electronic coupling $J_{ij}$ is a key factor which determines the delocalisation of the wavefunction in the crystal. In organic molecules this hopping term is anisotropic and therefore is linked to the anisotropy of the mobility. The determination of the values of $J_{ij}$ has been considered in the literature \cite{6Newton,6Bixon,6HSU,6fartini2009,6Ciuchi2011,Ciuchi2011}. The diffusion of charge carriers is also sensitive to several sources of  disorder that can be classified in two categories \cite{6simone2,6simone1,6Fratini2014}:
\vspace{0.5\baselineskip}
\begin{itemize}
\item \it Intrinsic disorder: \rm It's manifests through the variation of random values of the hopping integrals $J_{ij}$  for temperatures $T\neq 0$ K. This disorder comes from the displacement of the molecules under their thermal kinetic energy  at finite $T$. Because the Van der Waals interaction energy between molecules is relatively weak, the displacement of the molecules are large and the relative variation of the hopping integrals is also large with respect to their average values. 
\item \it Extrinsic disorder: \rm Impurities and defects are at the origin of this type of disorder.  The strength of the extrinsic disorder is often represented by a Gaussian distribution of the on-site energies  $\epsilon_i$ 
{
with the standard deviation $\Delta$.
}
\end{itemize}

\subsection{Modeling the hopping integrals}

\begin{figure}[h]
    \centering
    \includegraphics[width=0.6\linewidth]{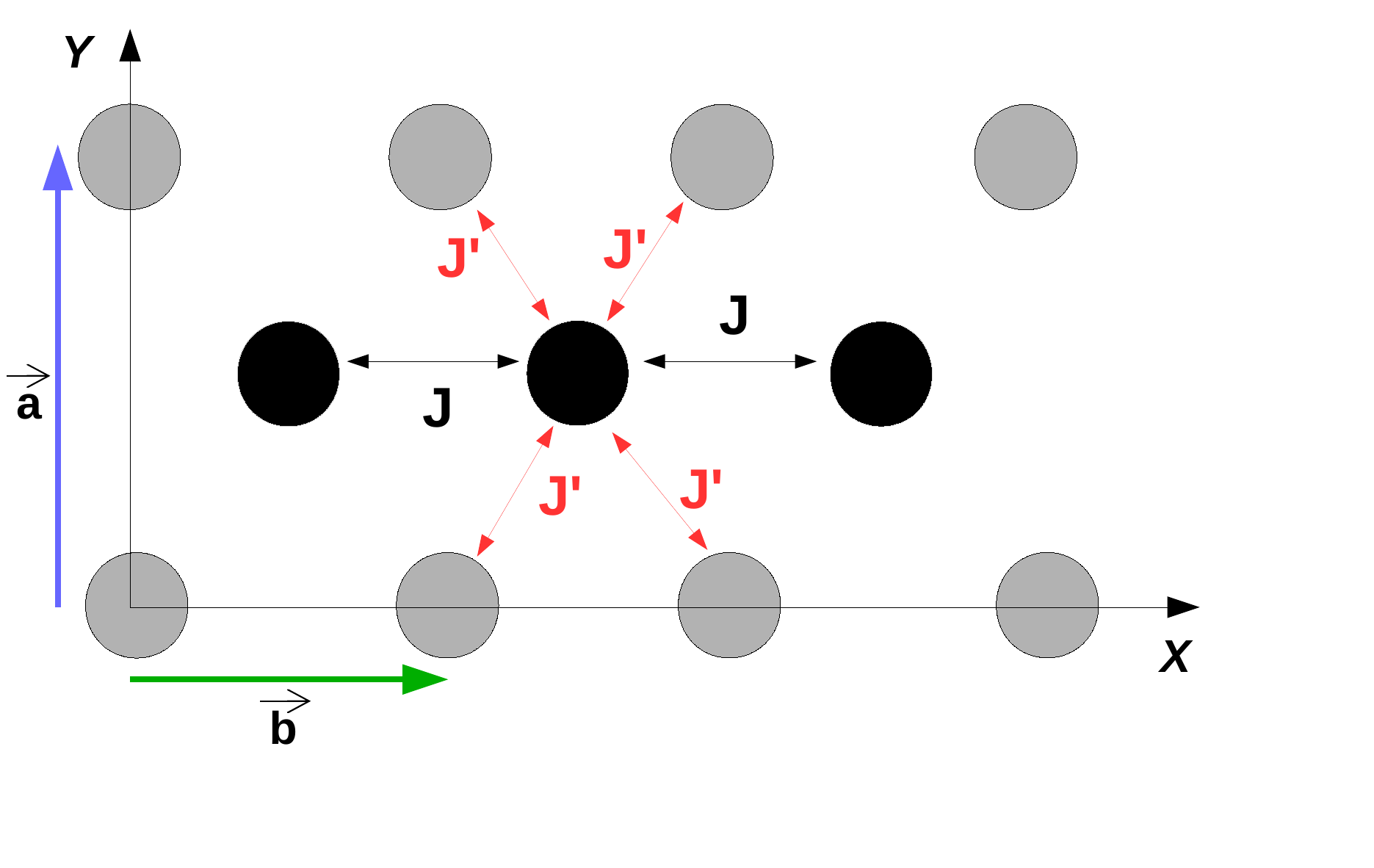}
     \caption{ 
{
2D model for nearest-neighbors couplings between rubrene molecules in the $ab$ layer \cite{6troisi2011}.} 
The $J$ coupling corresponds to the intra-chain coupling and has the highest value. The $J'$ coupling corresponds to the inter-chain coupling.  When $J'$ is neglected the rubrene model becomes that of decoupled chains (1D model). 
The vectors of the unit cell  $\vec{a}$ and $\vec{b}$ are of 1.44\,nm and 0.72\,nm respectively and are represented in the  $y$ and $x$ directions.}
     \label{2DD}
\end{figure}

In the tight-binding model one considers only the nearest-neighbor coupling \cite{6troisi2011}. At zero temperature, without disorder, the intra-chain integral is noted  $J$ and is of  $J=0.142$ eV. The inter-chain coupling  is noted  $J'$ with a value $J' = 0.028$\,eV \cite{6troisi2011}. This important difference implies an anisotropy which is indeed observed for mobilities in the plane along the directions $\vec{a}$ and $\vec{b}$ (figure \ref{2DD}).

\subsubsection{Correlated disorder}

\begin{figure}[h]
\centering
\includegraphics[width=0.57\linewidth]{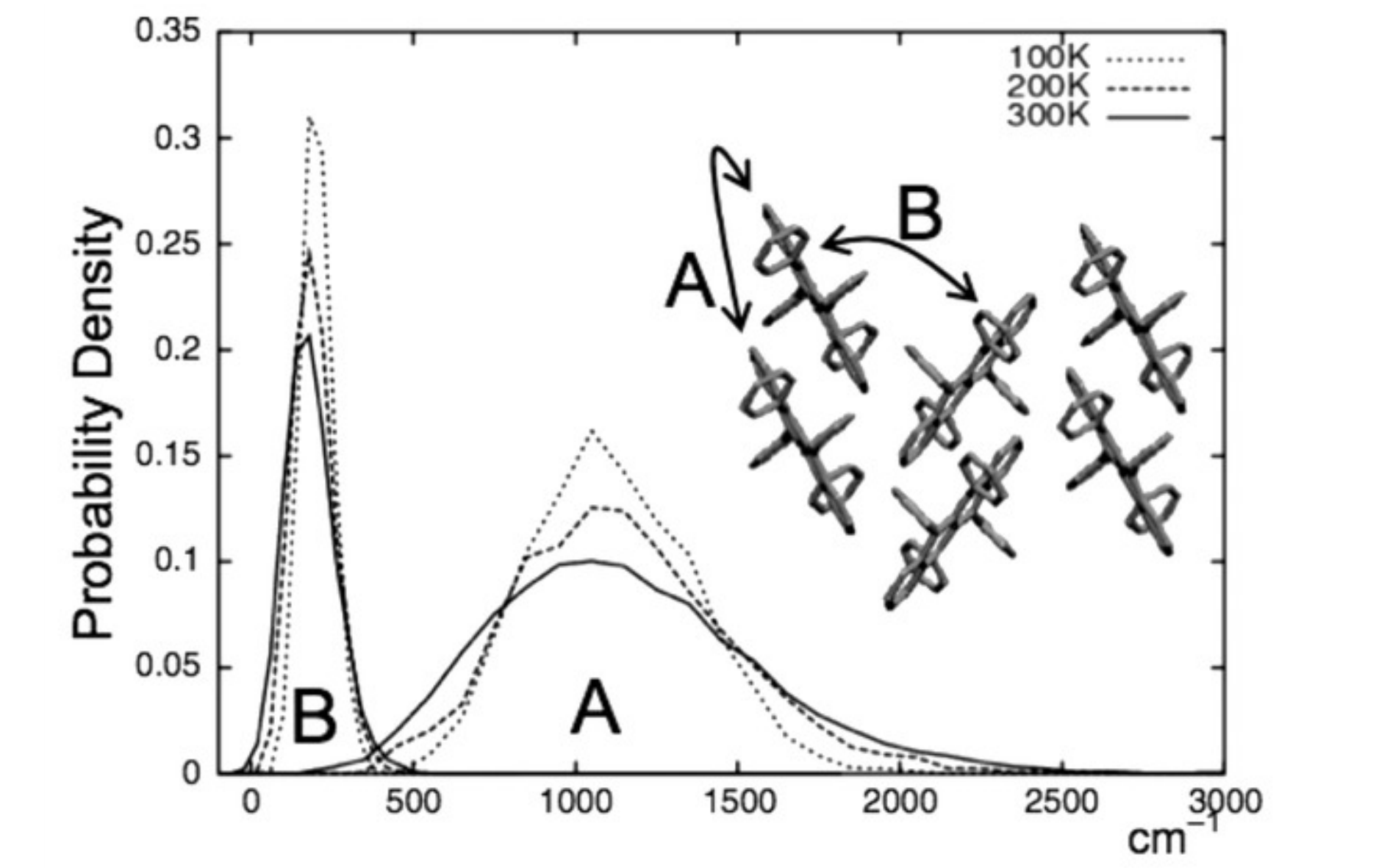}
 \caption{Probability distribution of hopping integrals for different molecular pairs corresponding to intra- and inter-chain hoppings
 {
 in rubrene,
   $A = J$ and $B = J'$ 
 }\cite{6Troisi2007}.\label{gauss}}
\end{figure}

In our calculations the vibrations of the molecules of rubrene around their equilibrium positions along the $(Oz)$ axis are taken into account for the hopping integral. These oscillations are equivalent to harmonic oscillators which are all identical, because the molecules are all identical and in equivalent positions. Their stiffness is given by $K$. The hopping integrals between molecular orbitals are extremely sensitive to small molecular displacements  \cite{6troisi}. It has been also  shown that the fluctuations are comparable to the mean-values \cite{6Troisi2005} (see figure \ref{gauss}).
The hopping integral is represented by Ref. \cite{6troisi},
\begin{equation}
J_{ij} = t(1-\alpha\Delta Z_{ij})\,,
\end{equation}
where  $t$ is the mean value at T = 0 K, $t = J$ or $J'$,  $\alpha=0.5831$ is the electron-phonon coupling \cite{6troisi} and  $\Delta Z_{ij} = \left | Z_{i}-Z_{j}\right |$, $Z_{i}$ and $Z_{j}$ are respectively the positions of the molecules  $i$ and $j$ along the axis (OZ). The probability for obtaining a given position $Z_{i}$ follows a Gaussian distribution,
\begin{equation}
G(Z_i) = \frac{1}{\sqrt{\beta K}}\, {\rm e}^{-\frac{\beta\, K\, }{2}Z_{i}^{2}} \,, \quad\beta=\frac{1}{k_B T} \,,
\end{equation}
where  $k_B$ is the Boltzmann constant , $T$ the temperature, $Z_{i}$ takes random values, 
{
and $K$ is stiffness of harmonic oscillators.
From  intermolecular vibration mode of pulsation $\omega_0 \simeq 5$-$9$\,meV \cite{6Troisi2007}, one can estimates $K = m \omega_0^2 \simeq 0.01\,{\rm eV\, nm^{-2}}$, with $m$ the molecular mass of rubrene.
}

If one considers three interacting molecules M$_{1}$,  M$_{2}$ and M$_{3}$ the coupling between M$_{1}$ and M$_{2}$ and the coupling between M$_{2}$ and M$_{3}$ are strongly correlated because, for example, the motion of M$_{2}$ alone changes both the coupling between  M$_{1}$-M$_{2}$ and between M$_{2}$-M$_{3}$. 

\subsubsection{Uncorrelated disorder}

Yet in the following we shall also consider a model (the uncorrelated model) where all the correlation between different hopping integrals is neglected \cite{Fratini17}. In that case, the hopping integrals are independent variables with a Gaussian distribution which reproduces the results in figure \ref{gauss}. In our work  fluctuations are represented by a Gaussian of variance $S$ given by \cite{6Fartini2016,6simone2}:
\begin{equation}\label{que}
S=\sqrt{4\lambda T k_{B} t},
{
~~{\rm with~} t = J ~{\rm or}~J'},
\end{equation}
where $T$ is the temperature,  the hopping integral is of $J=0.142$ eV for intra-chain (A type figure \ref{gauss})) and of  $J'=0.028$ eV for inter-chain (B type). $\lambda$ is the electron-phonon coupling. One has:
\begin{eqnarray}
\lambda = \frac{\alpha^{2} t}{2 M\omega_{0}^{2}}, 
{
~~{\rm with~} t = J ~{\rm or}~J'},
\end{eqnarray}
where $M$ represents the molecular mass and $\omega_{0}$ is the intermolecular vibration frequency. For rubrene  $\omega_{0}\simeq 4$-$9$\,meV \cite{6Troisi2007}. $\lambda$ is determined by fitting the curves in  figure \ref{gauss} with a Gaussian for both types of hopping. One get  $\lambda_{1}=0.15$ for intra-chain coupling and $\lambda_{2}=0.023$ for inter-chain coupling.

\section{Transient localisation model for charge transport in rubrene} 
\label{sec_transport}

{
We present now a study of electronic transport in rubrene in the transient localisation scenario \cite{6Ciuchi2011,6Fartini2016}, that seems so far the best model in the case of holes in rubrene. Here we consider only the case of holes because they have the highest mobility (compared to rubrene doped with electrons) and are the best candidates for the transient localization scenario. We present the effective Hamiltonian of holes. This means that the most relevant part of the spectra presented below are at the lowest energies since the thermal energy of holes is much smaller than their bandwidth.} In this scenario  there is not the formation of a polaronic state as explained above because the rubrene molecules are rigid and weakly distorted by the presence of an additional charge carrier. This absence of polaronic state is also  in accordance with the fact that the mobility is relatively high in hole doped rubrene. The transient localisation scenario  emphasizes the role of disorder that is due either to thermal motion or to impurities and chemical disorder. At short times, the disorder can be seen as static and Anderson-like localisation is produced. Yet at longer time scales,  the thermal disorder which is due to atomic displacements (phonons) of large amplitudes is dynamic and this variation with time of the potential landscape produces delocalisation and therefore allows mobility of charges.

\subsection{The Kubo formalism and the relaxation time approximation} 
\label{SecKubo}

Our starting point to compute the dc and ac conductivity is the Kubo formalism. This formalism has been applied with success to analyze the charge transport in many systems where the semi-classical model cannot be applied. This includes quasicrystals  \cite{Roche97,Triozon02,Trambly17}, organic semi-conductors \cite{6Fratini2014,Fratini17}, graphene and related systems 
\cite{Lherbier08,Lherbier11,Roche12,Trambly11,Trambly13,Trambly14,Soriano15,Missaoui17,Missaoui18}. In the case of  non-degenerated electron gas (with a Boltzmann occupation of states) the conductivity is given by \cite{6Fratini2014,6Ciuchi2011}:
\begin{equation}
\sigma\left(\omega\right)=-n e^{2}\omega^{2}\frac{\tanh{(\beta\hbar\omega/2)}}{\hbar\omega}\, {\rm Re}\int_{0}^{\infty}e^{i\omega t}\Delta X^{2}(t){\rm d}t \, ,
\label{opt}
\end{equation}
where $\beta=1/(\rm k_BT)$ where $k_B$ is the Boltzmann constant, $n$ is the charge carrier density, 
{
and $\Delta X^{2}(t)$ is the thermodynamic average 
}
of the mean-square displacement of a charge carrier along the chosen axis. 
We have developed powerful method \cite{Mayou88,Mayou95,Roche97,Roche99,Triozon02} that allows to compute $\Delta X^{2}$ efficiently and that have already been used for many systems  \cite{Triozon02,Trambly17,6Fratini2014,Fratini17,Lherbier08,Lherbier11,Roche12,Trambly11,Trambly13,Trambly14,Soriano15,Missaoui17,Missaoui18}.

The above expression is valid in the case of  a time independent Hamiltonian for the charge carriers. Yet in our situation the molecules move slowly, compared to typical electronic hopping times. For example in rubrene the intermolecular vibration frequency is of the order of a few  meV  ($\omega_0\simeq 6.2$ meV) whereas the typical hopping integral in the main direction is $J\simeq142$ meV \cite{6Ren2009,6Girlando2010,6Fartini2016}. These molecular displacements introduce a dynamical disorder on times of the order of  $\tau_{in}\simeq 1/\omega_0$. We show below a simplified treatment of the effect of this dynamical disorder in the context of a relaxation time approximation (RTA). This RTA is expressed from a central quantity which is the velocity correlation function $C(t)$,
Let us recall that the velocity correlation function $C(t)$ is related to the quadratic displacement $\Delta X^{2}(t)$ through the exact relation 
{
${\partial^2 \Delta X^{2}}/{\partial t^2}=C(t)$
}
\cite{6Fratini2014,6Ciuchi2011}.

For short times, $t\ll\tau_{in}$, the molecular lattice appears as static and the velocity correlation function is not modified as compared to that of the static model. Yet for large times,  $t\gg\tau_{in}$, we expect that the memory of the initial velocity is lost and the velocity correlation function $C(t)$  is cancelled. Therefore we expect that because of the dynamical disorder the velocity correlation function is killed in a characteristic time  $\tau_{in}$. This is represented by the phenomenological equation:
\begin{equation}\label{6crta}
C(t) = C_{0}(t){\rm e}^{-\frac{t}{\tau_{in}}} \, ,
\end{equation}  
where $C_{0}(t)$ is the correlation function for the static system which is directly related to the spreading computed.
In this RTA approximation the system is never completely localized because the dynamical disorder kills the Anderson localization, as expected. In fact the mobility  is given by in terms of the diffusivity $D$ by:
\begin{equation}\label{mobility}
\mu=\frac{eD}{k_{B}T} =\frac{e}{k_{B}T}  \frac{L^{2}(\tau_{in})}{2 \tau_{in}} \, ,
\end{equation}  
where $L^{2}(\tau_{in})$ is an average of the spreading $\Delta X^{2}(t)$ in the static structure over a characteristic time $\tau_{in}$. A detailed discussion of the meaning of the RTA approximation can be found in \cite{6Fratini2014,6Ciuchi2011}. 

Note that in our approach $\tau_{in}$ is not expected to depend much on the temperature. Yet when the temperature increases the static disorder also increases and therefore the term $L^{2}(\tau_{in})$ at the numerator of (\ref{mobility}) decreases. In addition the term $T$ at the denominator of (\ref{mobility})  increases. Therefore  the  formula (\ref{mobility}) predicts a decrease of the mobility when the temperature increase. This is consistent with the experimental results observed in rubrene and other organic semi-conductors (see also figures Ref. \cite{6Karl2003}).

\subsection{Electronic transport in rubrene: 1D model}

In a first step the rubrene crystal is considered as a collection of decoupled linear chains. The electron hopping occurs only  between molecules of the same chain with an average value $J=0.142$\,eV and one neglects the inter-chain hopping, $J'=0$, shown in the 2D representation (figure \ref{2DD}).
{
In this section, we present  results with correlated disorder which is more realistic \cite{6Fartini2016}. 
}

 \begin{figure}[h!]
 \centering
  \resizebox{0.45\textwidth}{!}{
   \includegraphics{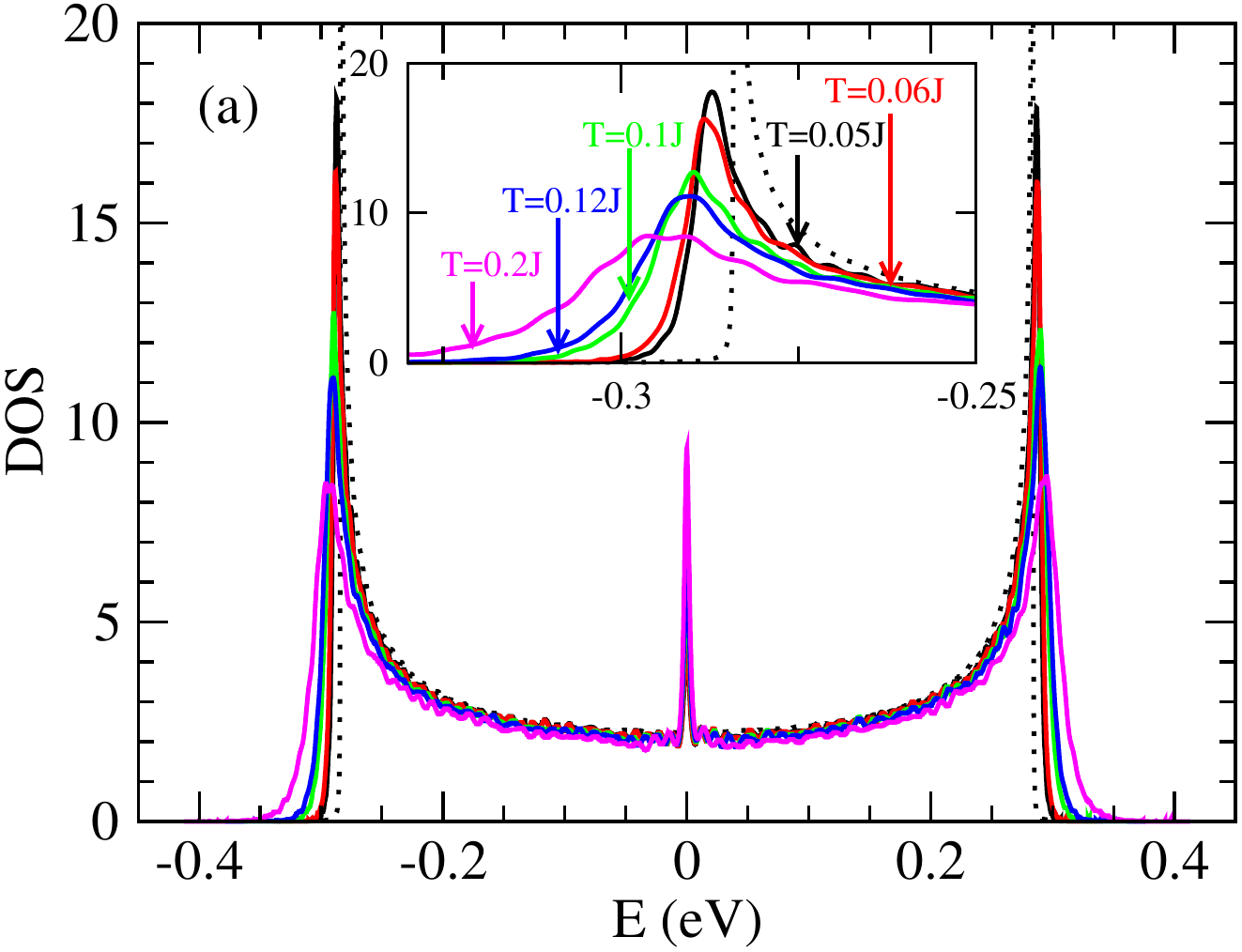}
   }
 \resizebox{0.45\textwidth}{!}{ 
   \includegraphics{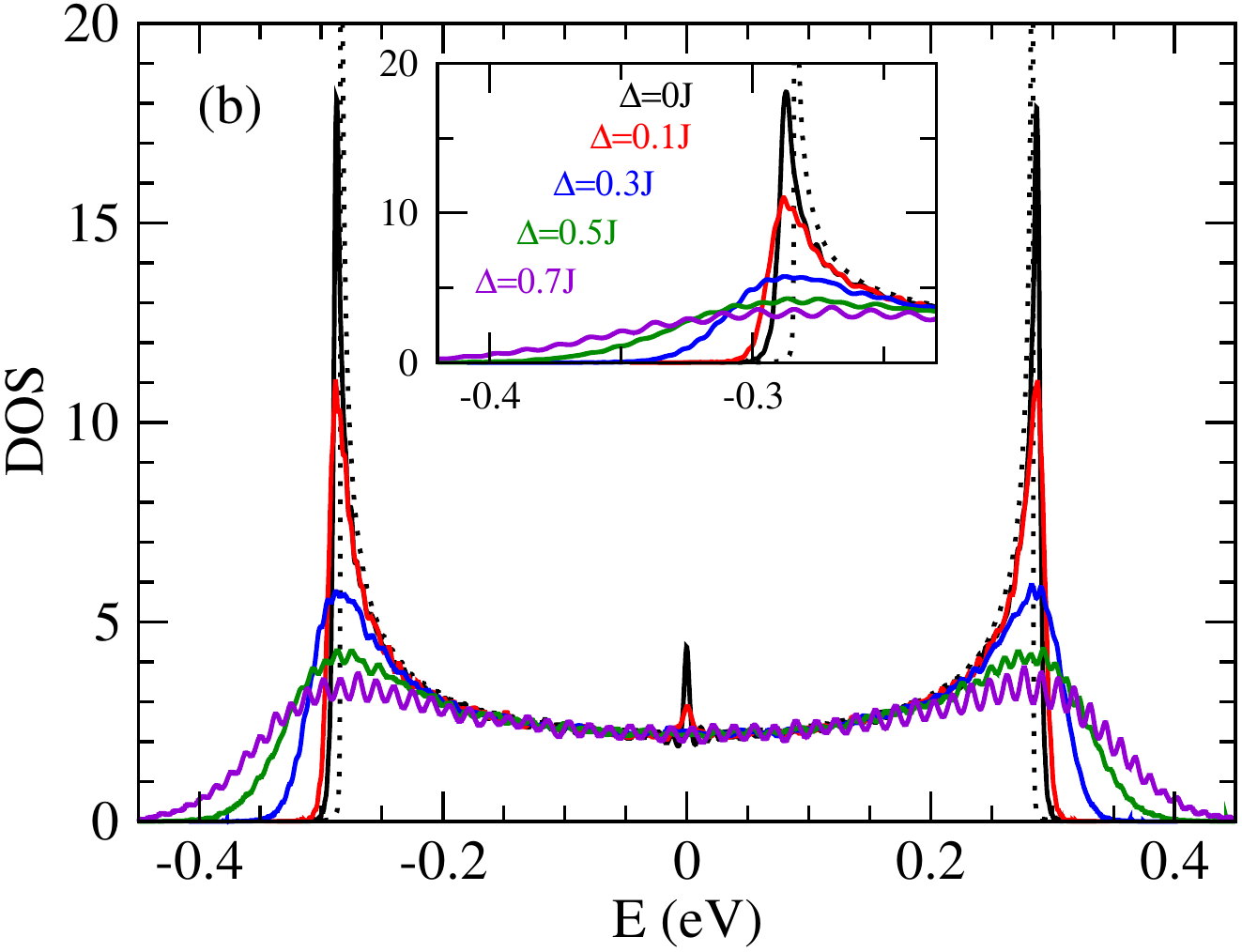}
 }
\caption{
{
Density of states (DOS) 
in the 1D model ($J'=0$) of rubrene: 
(a)
 with intrinsic intermolecular disorder for different temperatures and  $\Delta = 0$, 
(b) with intrinsic intermolecular disorder at temperature $T= 0.2 J$ and for different values of extrinsic disorder parameter $\Delta $. 
}
DOS unit is  $\rm states/(eV\,molecule)$ and $J=0.142$ eV.  \label{6dos1D} }
 \end{figure}
 
Figure \ref{6dos1D}(a) shows the evolution of the density of states (DOS) as a function of thermal disorder. This disorder is at the origin of the variations of the hopping integrals, in a random way, between adjacent molecules (let us recall that there is in the present model just one orbital per  molecule). The DOS without disorder (dotted line in figure \ref{6dos1D})  shows two Van Hove singularities at the band edges which are characteristics of one dimensional tight-binding models. The effect of the thermal disorder (intrinsic disorder) is to enlarge the DOS, with a visible effect in particular  at the band edges. There is the creation of a band tail of a few 100\,meV (figure \ref{6dos1D}(a)) which is controlled by the strength of the disorder. These states in the band tails are the states mainly occupied by holes (or electrons). This means that the effect of disorder for occupied states is particularly important. In the middle of the band, precisely at $E=0$,  there is a peak of DOS which existence is well known in one dimensional models with off diagonal disorder only \cite{6Eggarter1978,6Brouwer2000}. The diagonal disorder tends to destroy this structure as is evident in figure \ref{6dos1D}(b).


Figure \ref{6dos1D}(b) shows the effect of the extrinsic disorder due to defects through the parameter $\Delta$. The figure \ref{6dos1D}(b) shows the DOS for a temperature $T=0.2 J\simeq300$\,K and for different values of $\Delta$. The intrinsic singularities are destroyed either at the band edges or at the center of the band. The band tails become larger when the intrinsic disorder increases (i.e. when $\Delta$ increases).


\begin{figure}[!h]
\centering
\resizebox{0.7\textwidth}{!}{
  \includegraphics{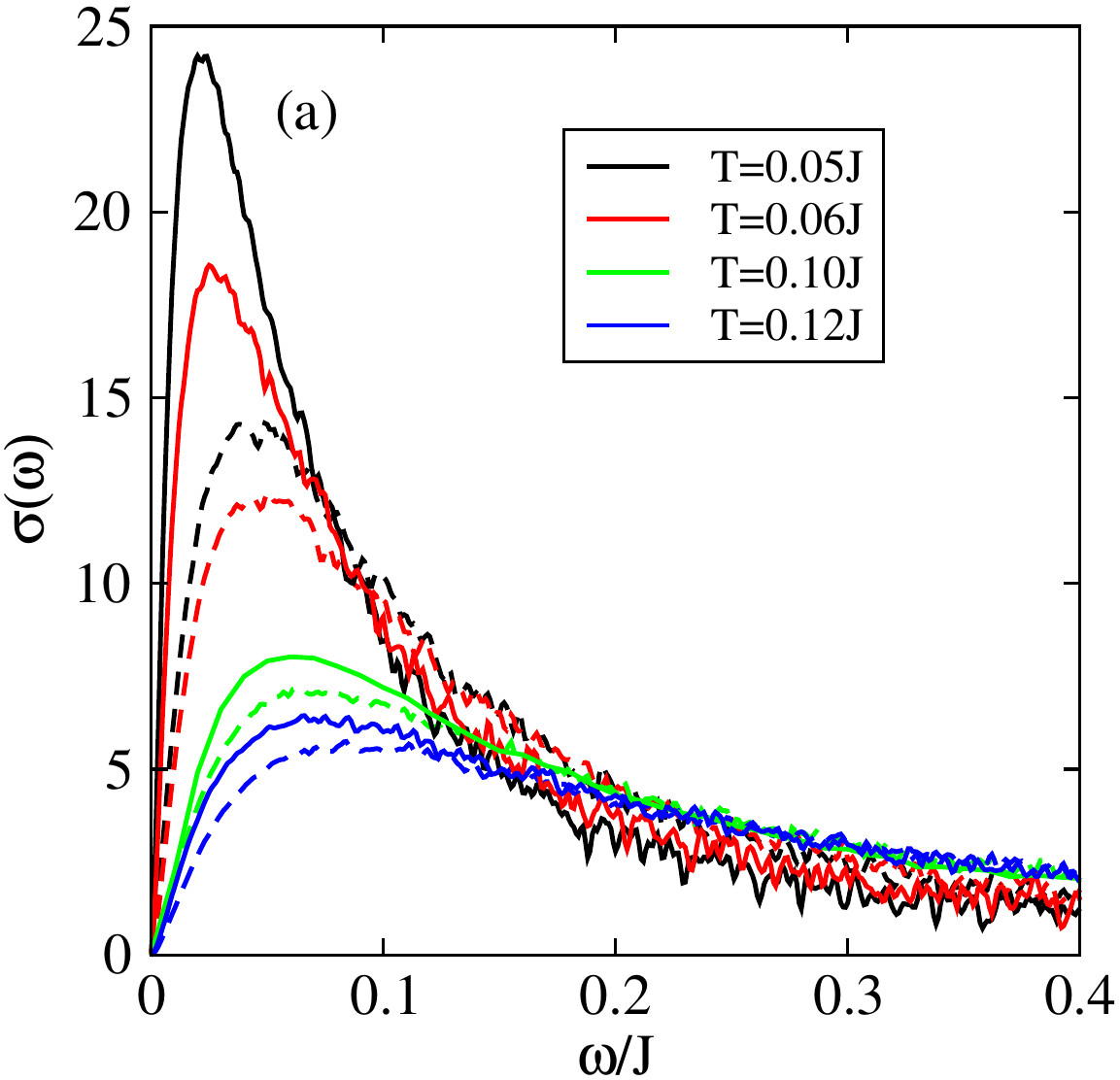}
   \includegraphics{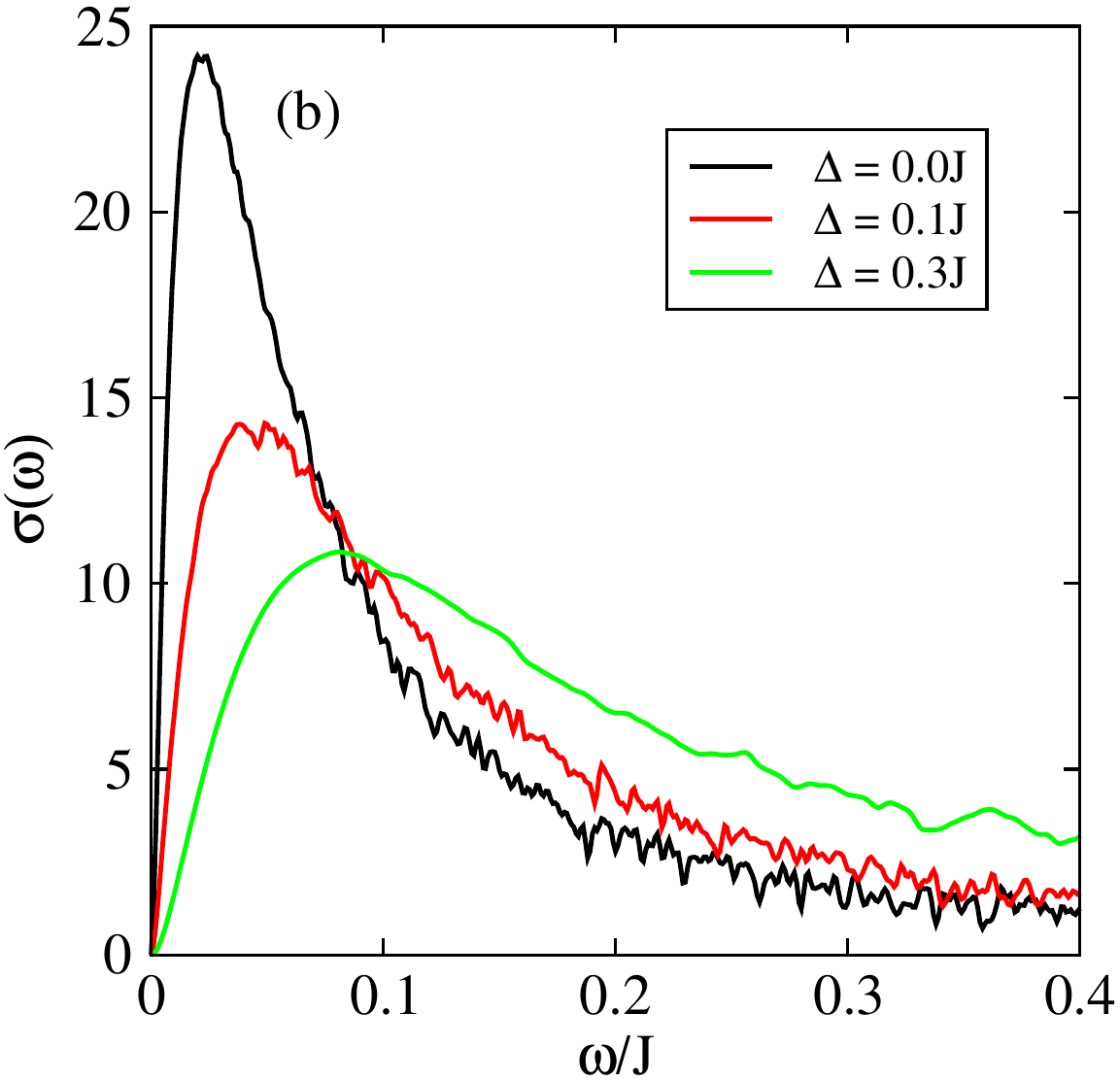}
  }
 \caption{Optical conductivity as a function of frequency in the 1D model ($J'=0$) of rubrene. (a) Different values of intrinsic disorder. Full line  $\Delta = 0$, dashed lines $\Delta = 0.1 J$. (b) Different values of extrinsic disorder $\Delta$ for a temperature  $T= 0.05 J$.
 $\sigma$ is given in unit of $\sigma_0=ne^{2}a^{2}/\hbar$ and $J=0.142$ eV. }
 \label{6condd}
\end{figure}

Figure $\ref{6condd}$ shows the optical conductivity, as a function of frequency, for different temperatures in the presence of intrinsic disorder. The results are given in units of $\sigma_{0} = ne^{2}a^{2}/\hbar$ where $a$ is the inter-atomic distance  and $n = N/V$ the density of charge carriers. When $\omega\rightarrow 0$, the conductivity tends to zero which corresponds to the fact that the systems is insulating when there is static disorder only in the 1D model (only states close to the band edges are occupied and the singular contribution of states close to energy $E=0$ for off diagonal disorder only can be discarded). The conductivity presents an absorption peak at a characteristic frequency which is governed by the disorder strength (figure \ref{6condd}). This peak is progressively suppressed and displaced towards higher frequencies when the intrinsic disorder increases \cite{6Fratini2014,6Uchida2013,6matthias2006} (figure \ref{6condd}(a)). This peak has been observed experimentally for an ambient temperature $T=0.18 J=300$\,K around the frequency $\omega=50$\,meV \cite{6matthias2006,6exper2,6Uchida2013}. We emphasize that this peak can be described only with models that treat quantum mechanically the transport of charge since this peak is intimately related to the cancellation of conductivity at zero frequency which is itself a manifestation of quantum localisation by disorder (Anderson transition).

\begin{figure}[h!]
\centering
 \resizebox{0.32\textwidth}{!}{
  \includegraphics{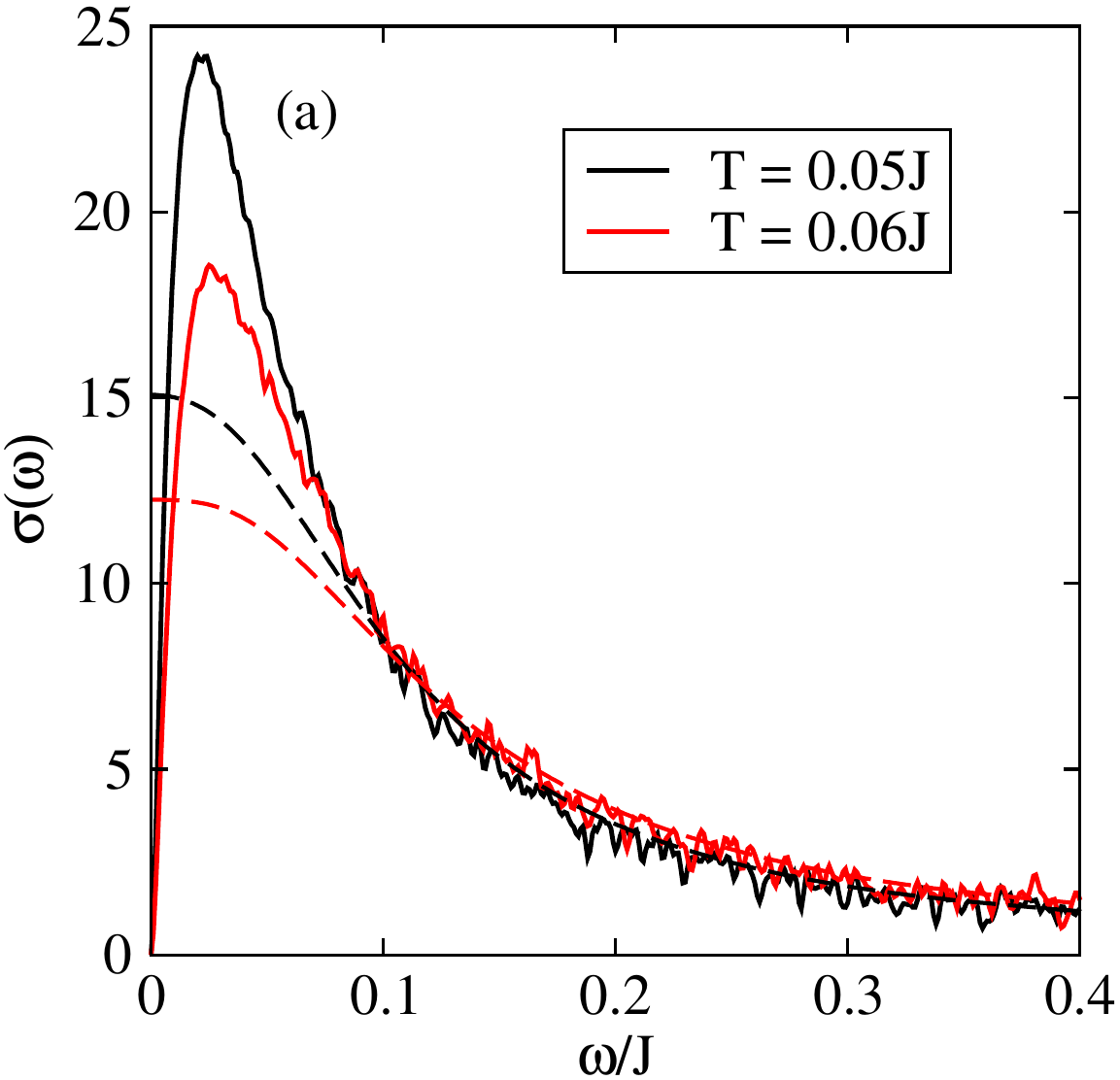}
  }
\resizebox{0.32\textwidth}{!}{ 
  \includegraphics{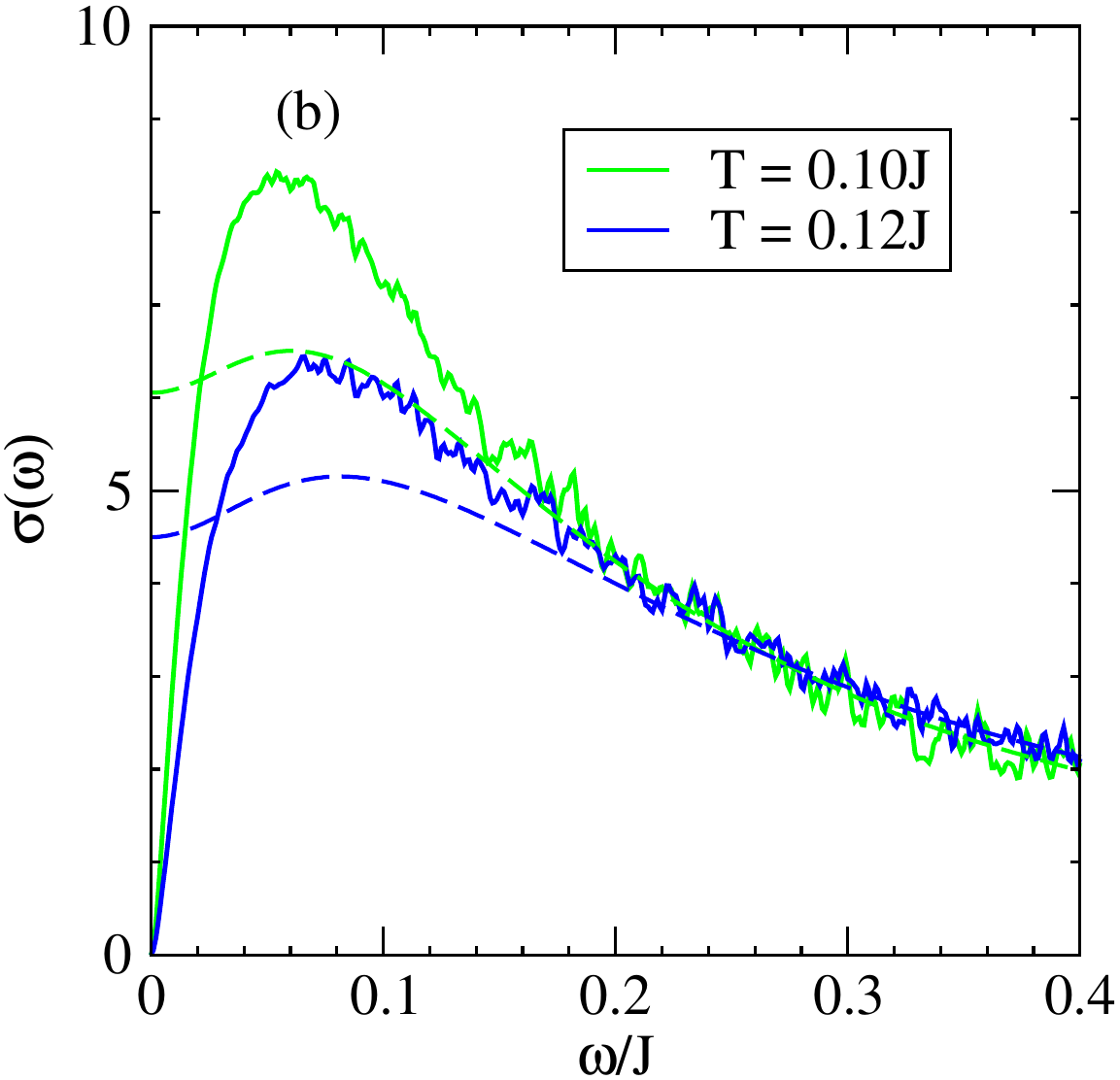}
}
\resizebox{0.32\textwidth}{!}{ 
  \includegraphics{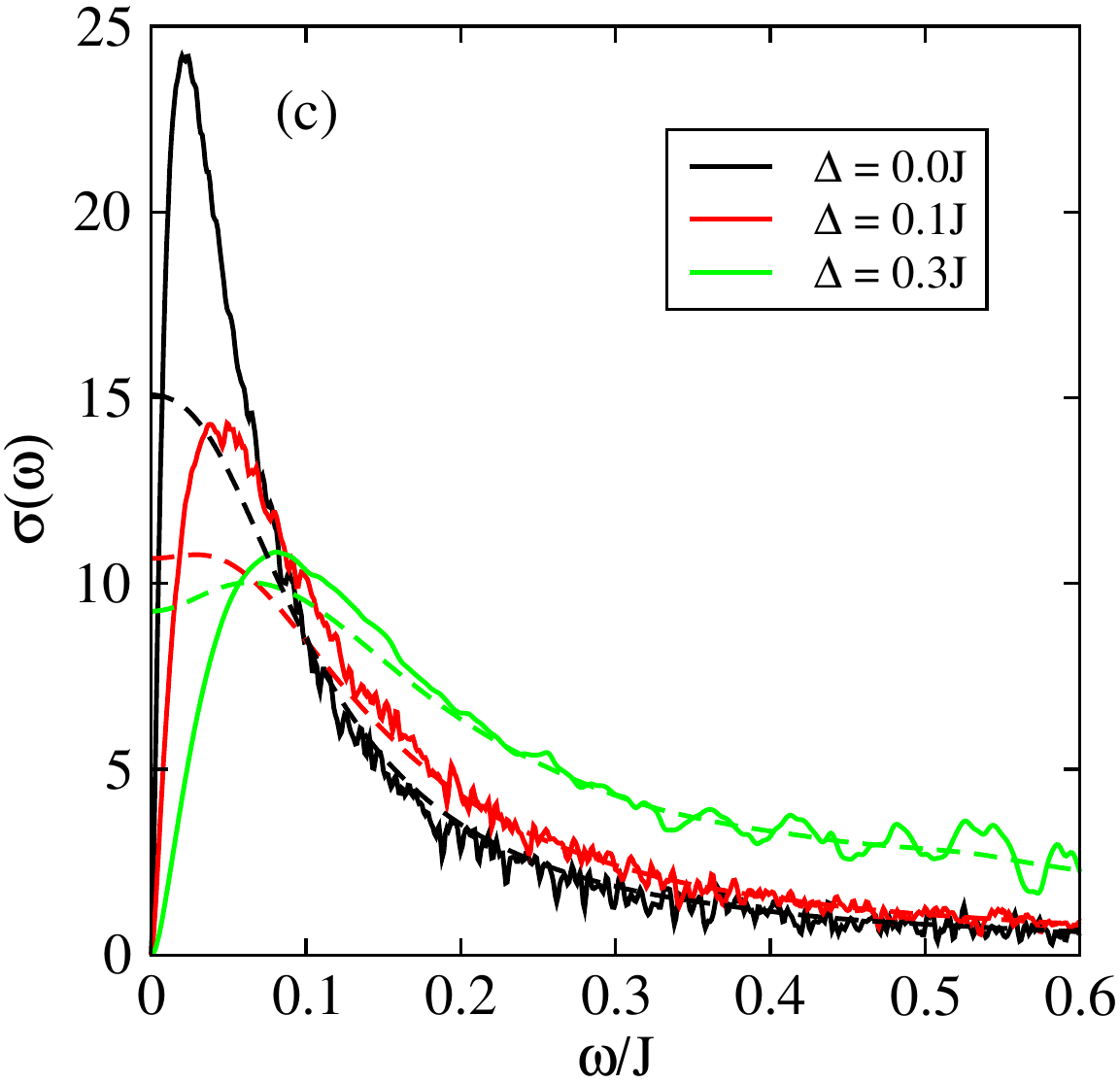}
}
 \caption{Optical conductivity as a function of frequency in the 1D model ($J'=0$) of rubrene: 
{
(full line) 
in the presence of static disorder, (dashed line) in the presence of static and dynamic disorder $\frac{\hbar}{\tau_{in}}=0.05 J$. 
(a) and (b) for different values of temperatures with $\Delta=0$, (c) for different values of  $\Delta$ for a temperature $T=0.05 J$. 
}
$\sigma$ is given in units of $\sigma_0=ne^{2}a^{2}/\hbar$ and $J=0.142$\,eV.\label{loc} }
 \end{figure}

The effect of the disorder coming from impurities and defects, which is represented by the extrinsic disorder parameter $\Delta$ \cite{6simone1,6simone2}, is shown in figure \ref{6condd}(b). The localisation peak appears at a value $\omega^{*}\simeq 0.05J$ for $\Delta = 0.1J$. This frequency depends on the value of $\Delta$ as shown in figure \ref{6condd}(b) for a fixed temperature. The value of  $\omega^{*}$ depends on the intensity and on the nature of disorder present in the system.

\begin{figure}[h!]
\centering
 \resizebox{0.4\textwidth}{!}{
  \includegraphics{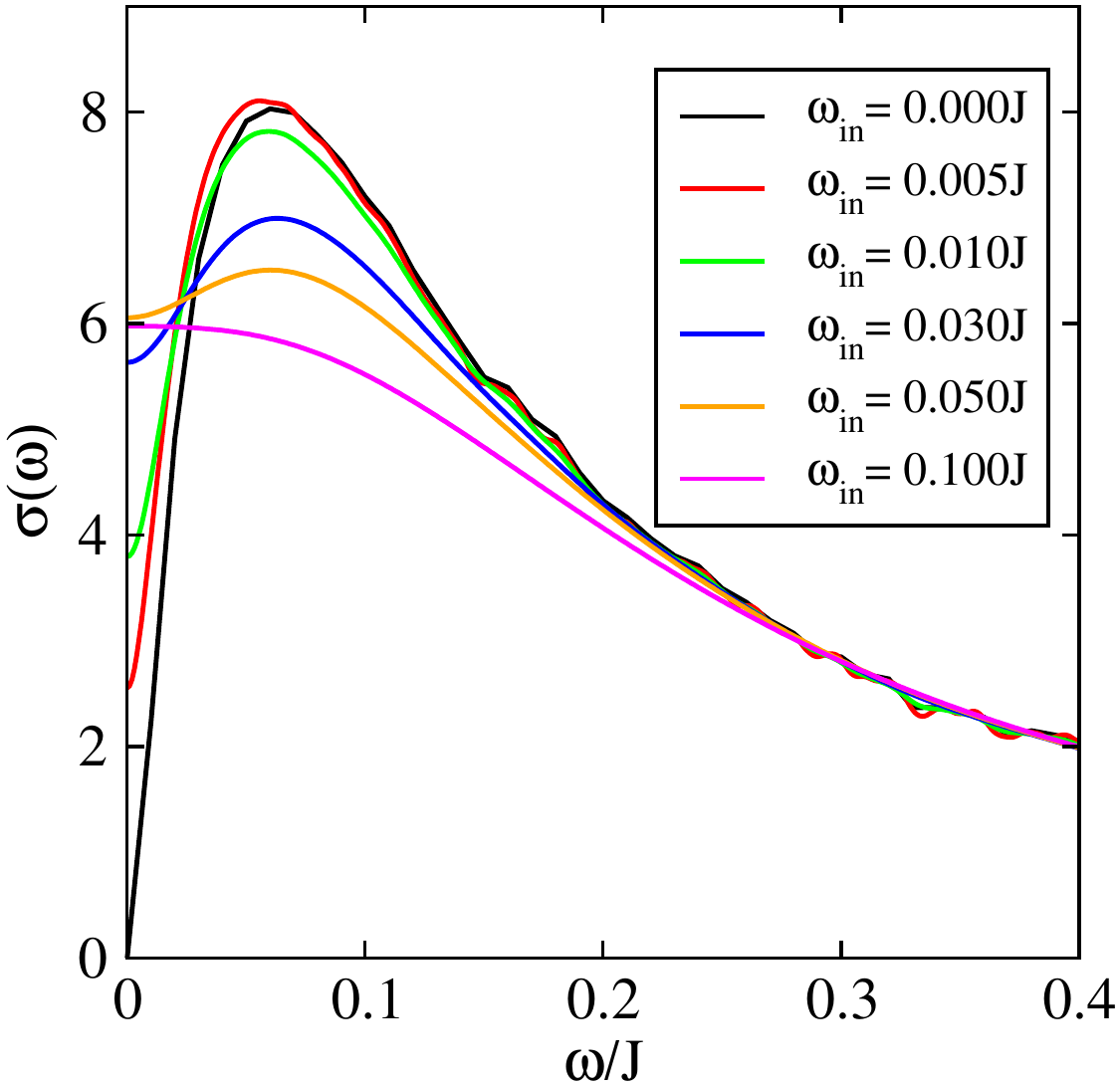}
  }
 \caption{Optical conductivity as a function of frequency in the 1D model ($J'=0$) of rubrene for a temperature $T=0.1 J$ and for different values of dynamic disorder  ($\omega_{in}\simeq\frac{1}{\tau_{in}}$). $\sigma$ is given in units of $\sigma_0=ne^{2}a^{2}/\hbar$ and $J=0.142$\,eV.  }\label{6pic}
\end{figure}

We discuss now the effect of dynamical disorder through the results presented in figures $\ref{loc}$ and $\ref{6pic}$. We apply the theory of dynamical disorder based on the relaxation time approximation applied to the velocity correlation function (see above section \ref{SecKubo}). Because of the loss of coherence of the scattering after a characteristic time $\tau_{in}$  Anderson localisation cannot be complete and the zero frequency conductivity is now positive. For dynamical disorder  $\hbar/\tau_{in}=0.05J$ (dashed lines in figures \ref{loc}(a) and (b)) the localisation peak has disappeared for the lowest temperatures i.e. for the lowest disorder. By contrast a localisation peak in the frequency dependent conductivity is still present for stronger disorder i.e. for higher temperatures (when there is no extrinsic disorder). An analogous behavior is observed if there is also a source of extrinsic disorder.


The value of dynamical disorder is determined by the intermolecular displacements as a function of time. On figure \ref{6pic}, the conductivity is presented as a function of frequency for different values of the inelastic scattering $\tau_{in}$. For $\hbar/\tau_{in}=0$ one is in the static model and the Anderson localisation operates fully (the dc-conductivity is zero). Yet, $\sigma(\omega=0)$ increases with  $\hbar/\tau_{in}$ as expected. $\sigma(\omega=0)$ reaches its maximum when $\tau_{in}$ is of the order of the time needed to reach the diffusive regime (elastic scattering time).
The semi-classical models allow to treat the optical conductivity at higher frequencies $\omega_{in}$ where localisation effects do not play a role. This is shown in figure \ref{6pic} by the convergence of the different curves (for different $\tau_{in}$) at high frequencies. 
{
This convergence at high frequency comes from the fact that the conductivity at high frequency is directly related to the quantum diffusion at small times (see equation (\ref{opt})). The diffusion at small times  is essentially ballistic and then diffusive and independent of the inelastic scattering time or even of the presence of localization effects.}  

\subsection{Electronic transport in rubrene: 2D models}
\label{Sec_result2D}

Since the interaction between chains in the rubrene crystal is weak most of theoretical studies \cite{6simone2,6troisi,6Fratini2014} have neglected this coupling (coupling terms $J'=0$ in figure \ref{2DD}). This assumption limits of course the theoretical studies to transport along the chains and gives no indication on the transport perpendicular to the chains and on the anisotropy of transport. Here we want to propose a study that allows a comparison with experiments concerning transport in any direction and therefore we include now the coupling between the chains. 

Because the description of disorder is important in the model of organic semi-conductors we take this opportunity to compare two microscopic models of disorder. On one side we consider  correlated disorder where the positions of the molecules are given according to the energy functional described above. In this model the displacement of the different molecules are uncorrelated because each molecules moves in its own potential well but the hopping integrals present a correlation because they depend on the same set of positions of the molecules. On the other side we consider a model with uncorrelated disorder for intrinsic disorder where each hopping integral between any two adjacent molecules is considered as an independent variable. Of course in this model the distribution of the values of the hopping integrals presents the same distribution as in the correlated disorder model.

\subsubsection{Correlated disorder}

\begin{figure}[h!]
\centering
 \resizebox{0.48\textwidth}{!}{
  \includegraphics{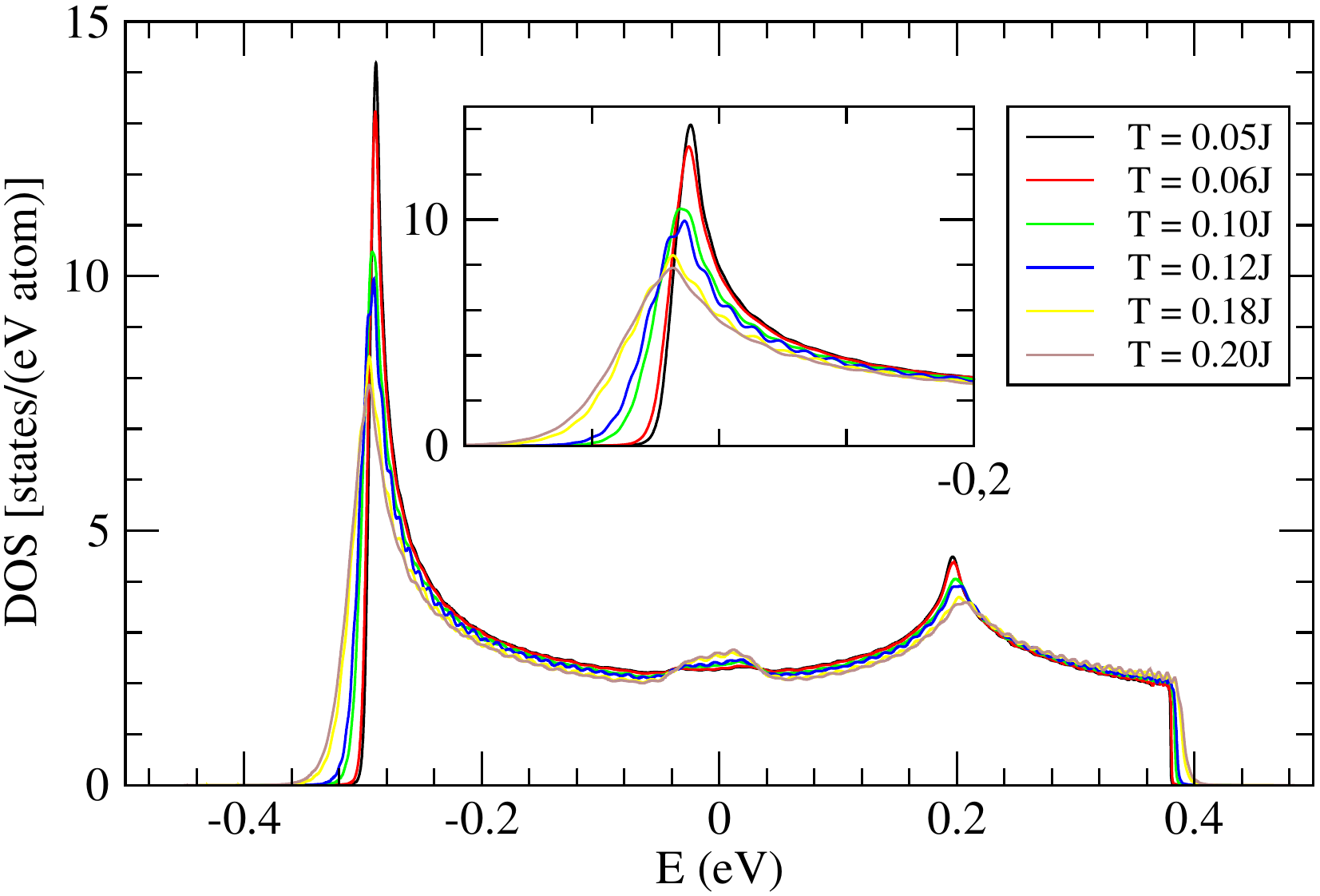}
  }
\resizebox{0.48\textwidth}{!}{ 
  \includegraphics{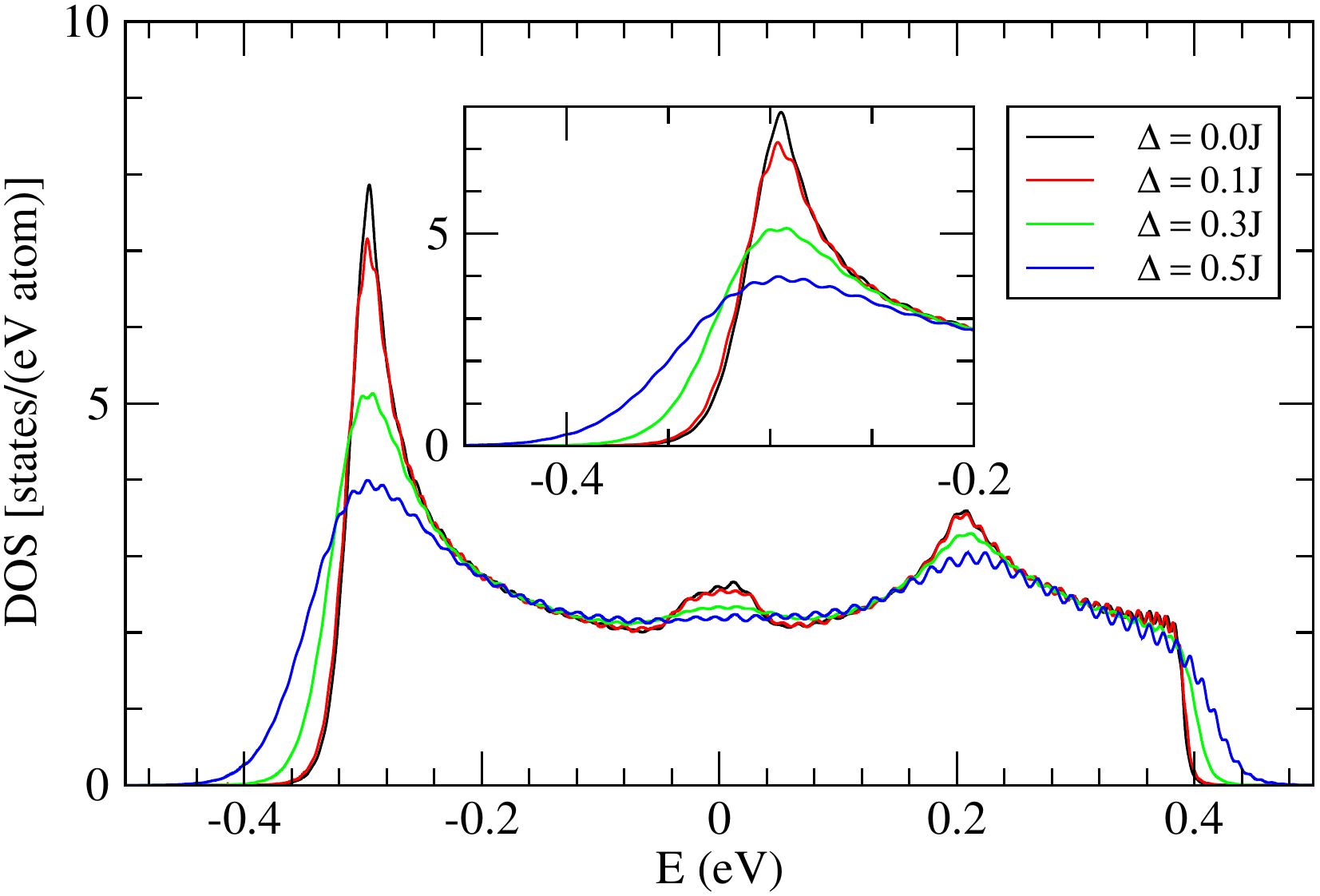}
}
\caption{Density of states (DOS) of a rubrene crystal in the 2D model: (a) Effect of intrinsic disordered ($\Delta=0$), (b)   Effect of extrinsic disorder for $T=0.2J$.
DOS in unit of  $[\rm states/(eV molecule)]$. $J=0.142$\,eV and $J' = 0.028$\,eV.
\label{6fig:conductivity_and0}  }
\end{figure}

Contrary to the 1D model the DOS in the 2D model is asymmetric with respect to the center of the band  (figure \ref{6fig:conductivity_and0}). This asymmetry is due to the weak inter-chain coupling. Note that the holes are in the band tails at low energy. The two types of disorder (intrinsic and extrinsic) have similar effect as in the 1D model.

\begin{figure}[h!]
\centering
\includegraphics[width=0.48\linewidth]{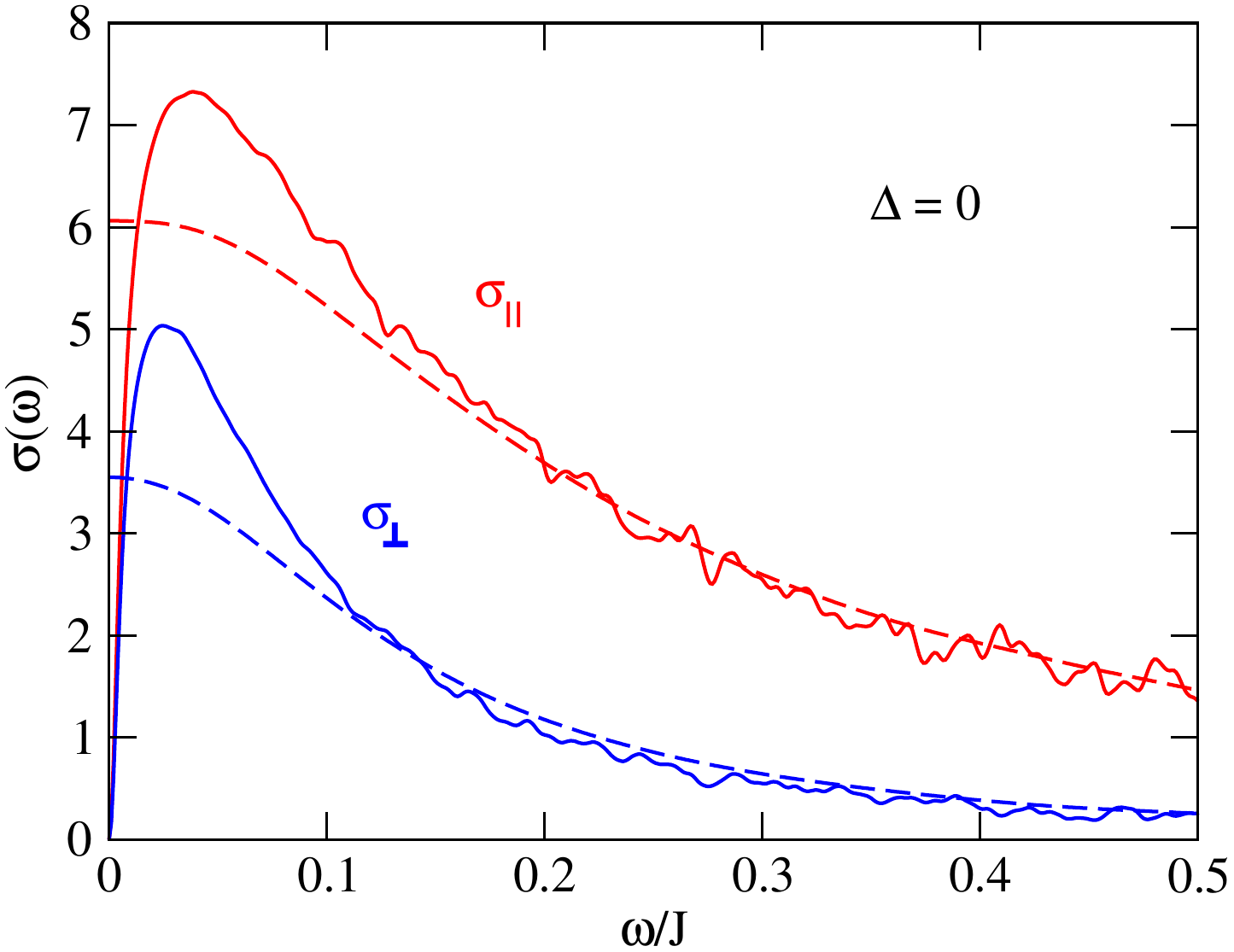}
\includegraphics[width=0.48\linewidth]{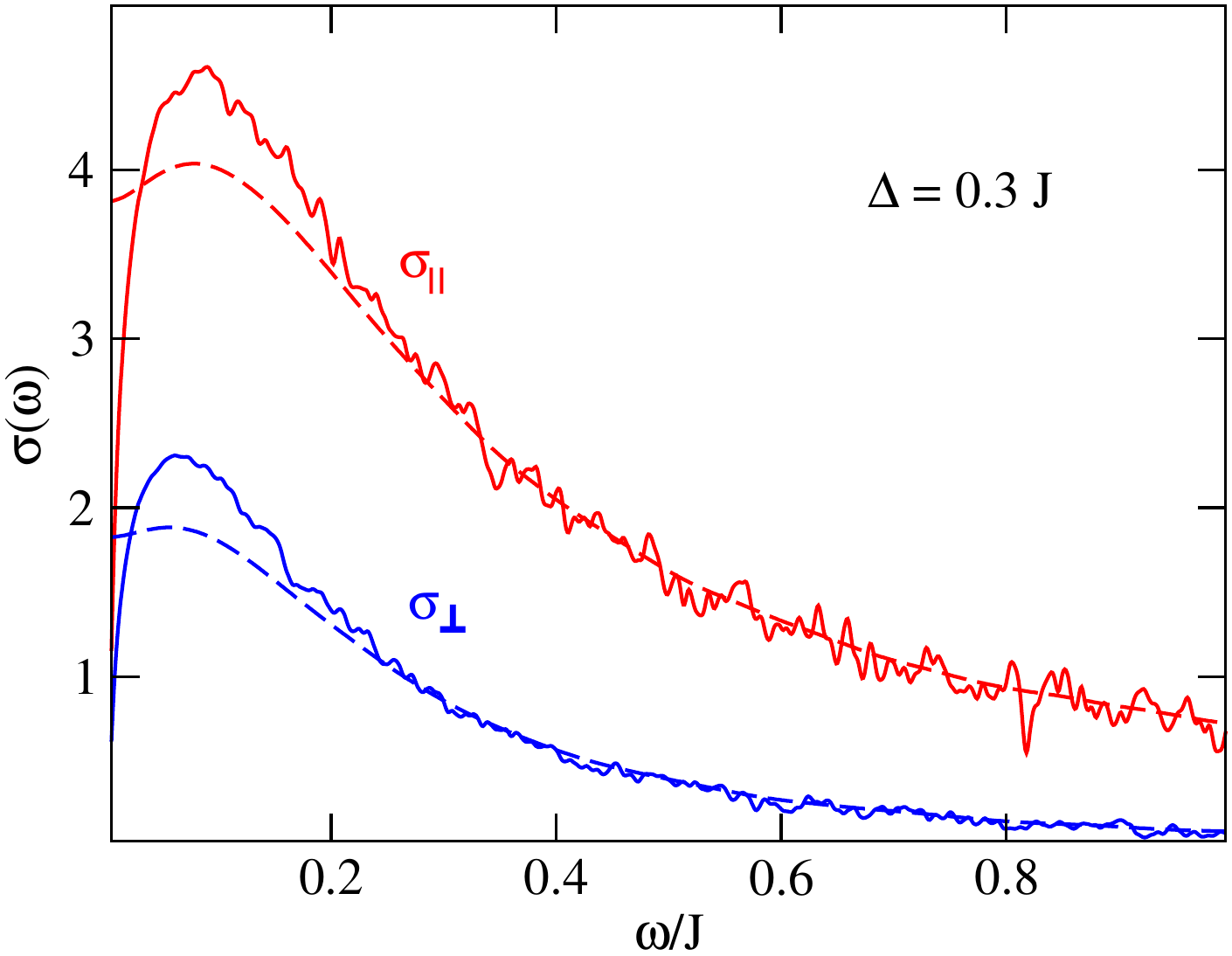}
\caption{Optical conductivity of rubrene in the 2D model in the presence of static disorder $T=0.18 J$ and dynamical disorder $\frac{\hbar}{\tau_{in}}\simeq0.05 J$: 
(Full line) static disorder only, (Dashed line) static disorder and dynamical disorder.
(a) Extrinsic disorder  $\Delta=0$, (b) Extrinsic disorder $\Delta=0.3 J$. 
$\sigma$ is given in units of $\sigma_0=ne^{2}a^{2}/\hbar$,  $J=0.142$\,eV and $J' = 0.028$\,eV.  }
\label{6fig:conduc_T018_and0}
\end{figure}
\paragraph{}

In the 2D model one can determine the optical conductivity perpendicular to the chains  (figure \ref{6fig:conduc_T018_and0}) $\sigma_{\perp}$ (conductivity in the direction $\vec{a}$ ) and the conductivity along the chains (direction $\vec{b}$) $\sigma_{\parallel}$. The localisation peak in the $2D$ model is present around the frequency $\omega^{*}\backsimeq40$ meV  in both directions $\vec{b} (\parallel)$ and $\vec{a} (\perp)$ (figure \ref{6fig:conduc_T018_and0}(a)) for an intrinsic disorder $T=0.18J$ and without extrinsic disorder $\Delta=0$. For a system with extrinsic disorder $\Delta=0.3J$ (figure \ref{6fig:conduc_T018_and0}(b)) $\omega^{*}\backsimeq70$\,meV. This slight difference with the experimental results ($\omega^{*}\backsimeq50$\,meV) is probably due to the estimation of the parameter $\Delta$ which describes the presence of defects and impurities which varies from one sample to an other. The anisotropy of the system is well demonstrated by the behavior of the conductivity in both directions. The perpendicular conductivity, $\sigma_{\perp}(\omega)$, is $\sim2$ times smaller than the parallel conductivity, $\sigma_{\parallel}(\omega)$. Despite the large difference between the inter and intra molecular hopping integrals (a factor of $\sim5$) the effect on the conductivity is relatively reduced.

\subsubsection{Uncorrelated disorder}
\begin{figure}[h!]
\centering
\includegraphics[width=0.48\linewidth]{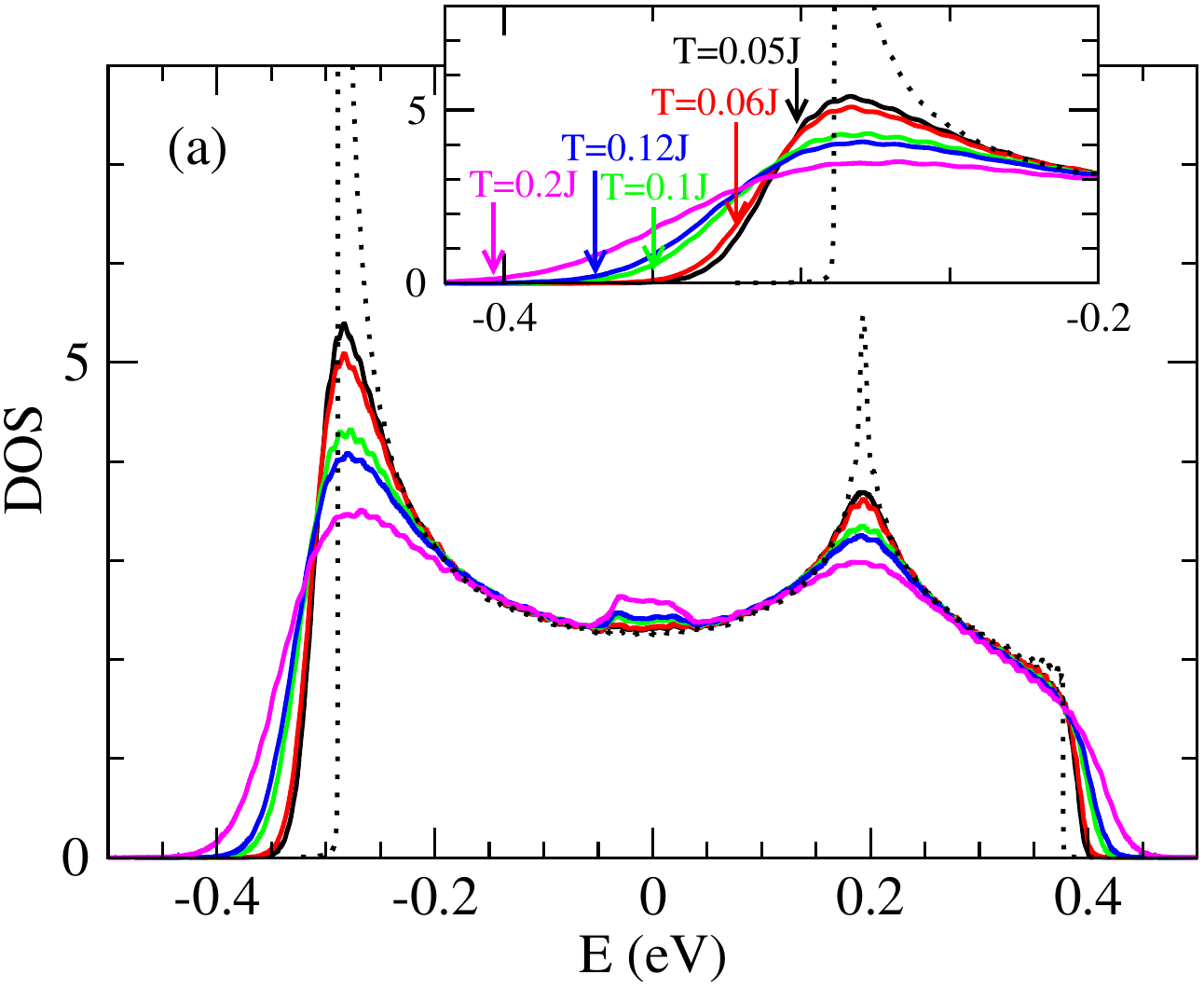}
\includegraphics[width=0.48\linewidth]{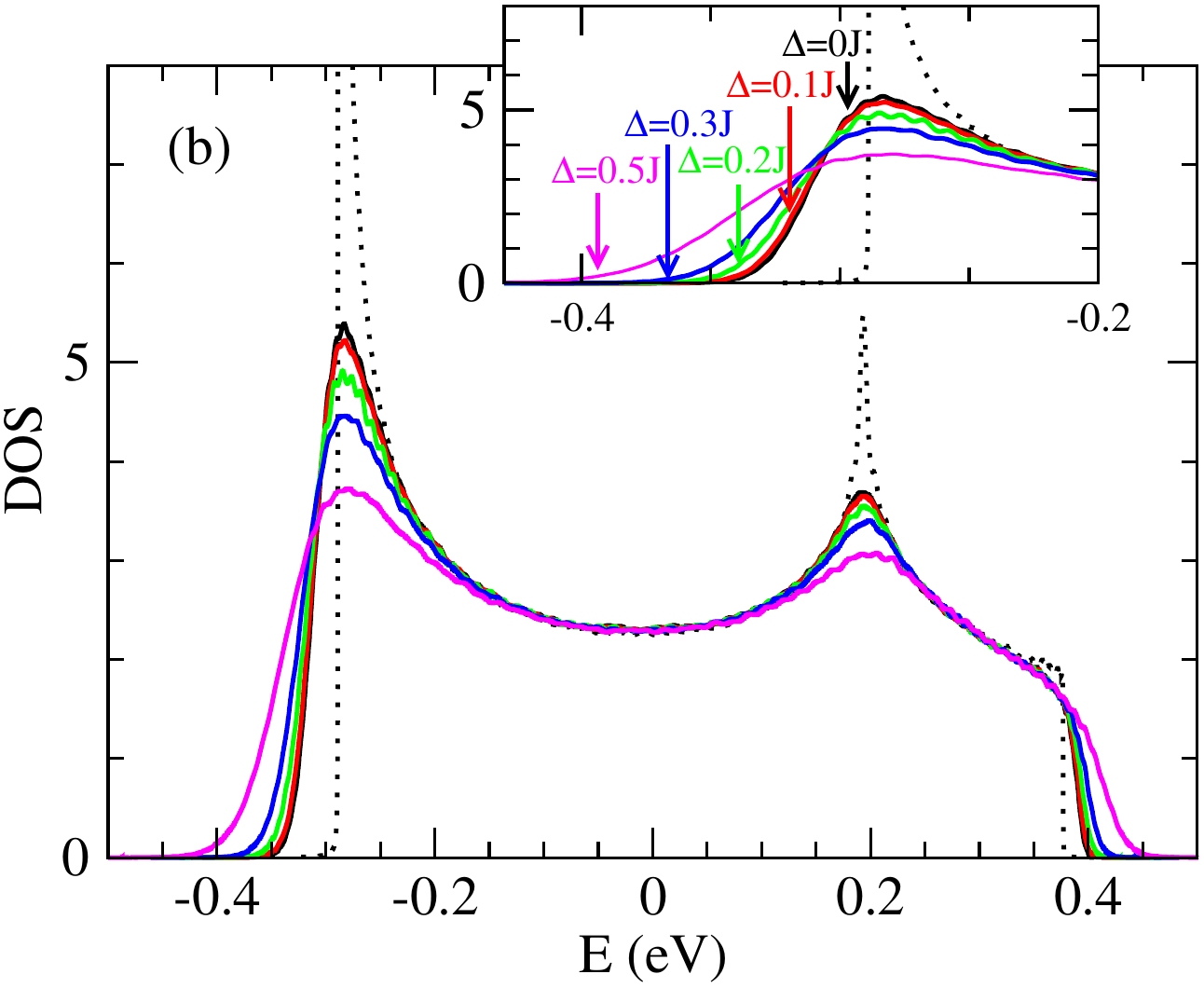}
\caption{Density of states (DOS) of rubrene in the 2D model with uncorrelated disorder.  (a) Only the intrinsic disorder is taken into account  (the values of the temperatures are indicated in the inset), (b) Effect of extrinsic disorder  $\Delta$ for a temperature $T=0.05 J$. The DOS without disorder is represented by the dotted line. $J=0.142$\,meV and $J' = 0.028$\,eV. DOS in unit of  $[\rm states/(eV molecule)]$.} 
\label{fig:DOS_2D_and0}
\end{figure}

In this part we present some results relative to the model of uncorrelated hopping integrals in the 2D model. The densities of states are represented respectively in figure \ref{fig:DOS_2D_and0}.

As in the correlated case the presence of intrinsic disorder has two remarkable effects on the DOS (figure \ref{fig:DOS_2D_and0}(a)). 
The first one is a rounding of the band tail which shifts toward higher energies, which indicates a broadening of the band. The second effect is the appearance of new states in the band tail which tend to be more localized states than in the band center. These two effects are controlled by the quantity of intermolecular fluctuations $S$ (equation (\ref{que})) of the hopping integral which increases with temperature.

One notices also that there is a peak of DOS close to the band center. 
{
This peak is relatively more important in the 1D case (figure  \ref{6dos1D})  than in the  2D case (figures \ref{6fig:conductivity_and0} and \ref{fig:DOS_2D_and0}). 
But it has no effect on the conductivity of holes that comes from states with the lowest energies of the bands.
In 2D case,
comparing the densities in the correlated disorder (figure \ref{6fig:conductivity_and0}) and uncorrelated disorder (figure \ref{fig:DOS_2D_and0}) 
}
at the same temperature the uncorrelated disorder affects more the DOS. If one looks for example the case of temperature $T=0.05 J$ in the  1D model the presence of correlated disorder lowers the two Van hove singularities at the band edges (figure \ref{6dos1D}(a)) whereas the uncorrelated disorder destroys completely these singularities (figure \ref{fig:DOS_2D_and0}).

\begin{figure}[h!]
\centering
\includegraphics[width=0.5\linewidth]{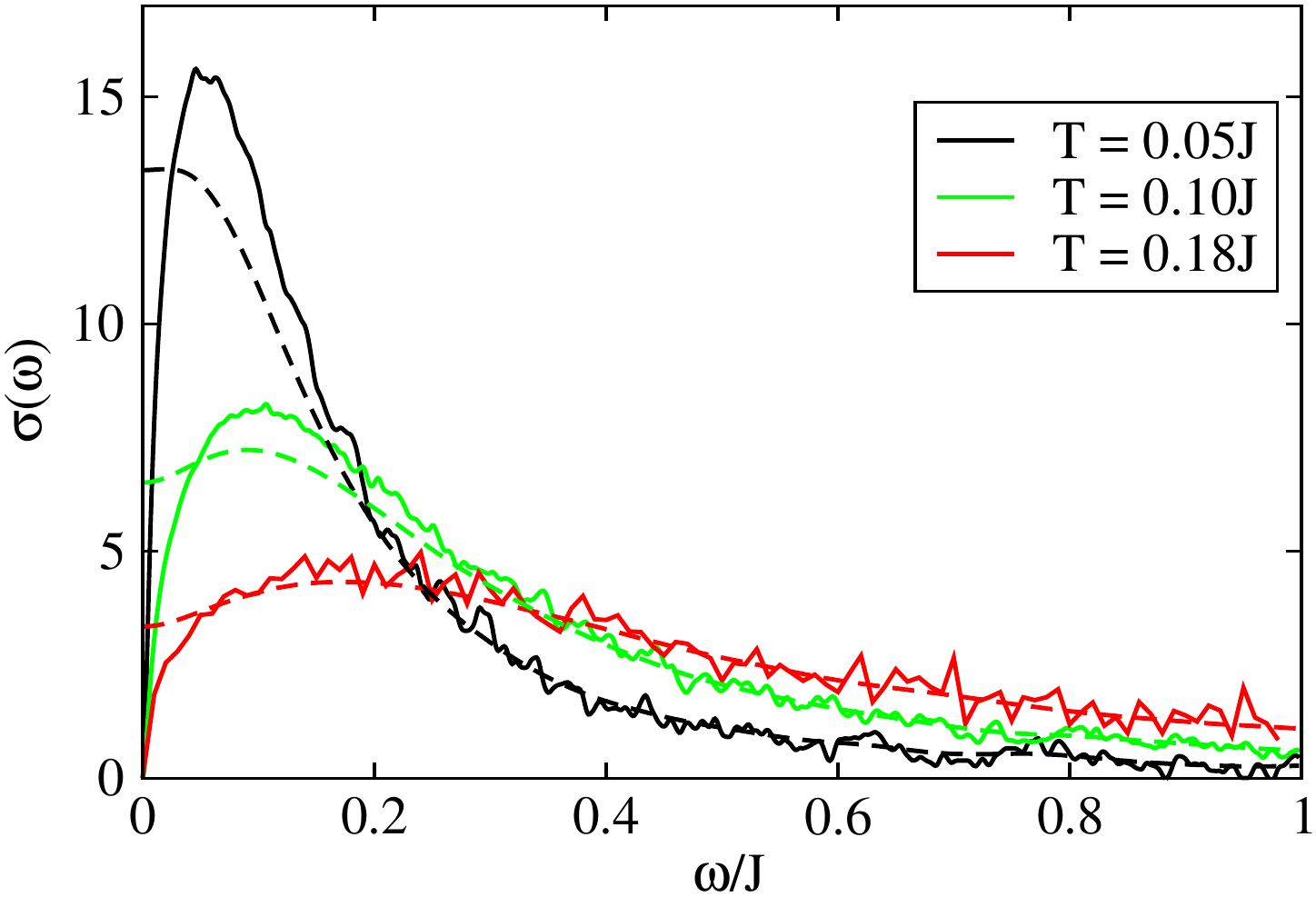}
\caption{Optical conductivity  $\sigma(\omega)$ of the  2D model with uncorrelated disorder for different temperatures in the absence of extrinsic disorder: 
(Full line) static disorder only, (Dashed line) static disorder and dynamical disorder with  $\hbar/\tau_{in}=0.05 J$. 
 $\sigma$ is given in unit of $\sigma_0=ne^{2}a^{2}/\hbar$, $J=0.142$\,eV and $J' = 0.028$\,eV.  }
\label{fig:cond_2D_non_and0}
\end{figure}

As noted previously the effect of extrinsic disorder (parametrized by $\Delta$) is to generate band tails, and to destroy the Van Hove singularities at the band edges as well as  the peak of less localized states at the center of the band close to $E=0$ eV (figure \ref{fig:DOS_2D_and0}(b)).

\begin{figure}[h!]
\centering
\includegraphics[width=0.47\linewidth]{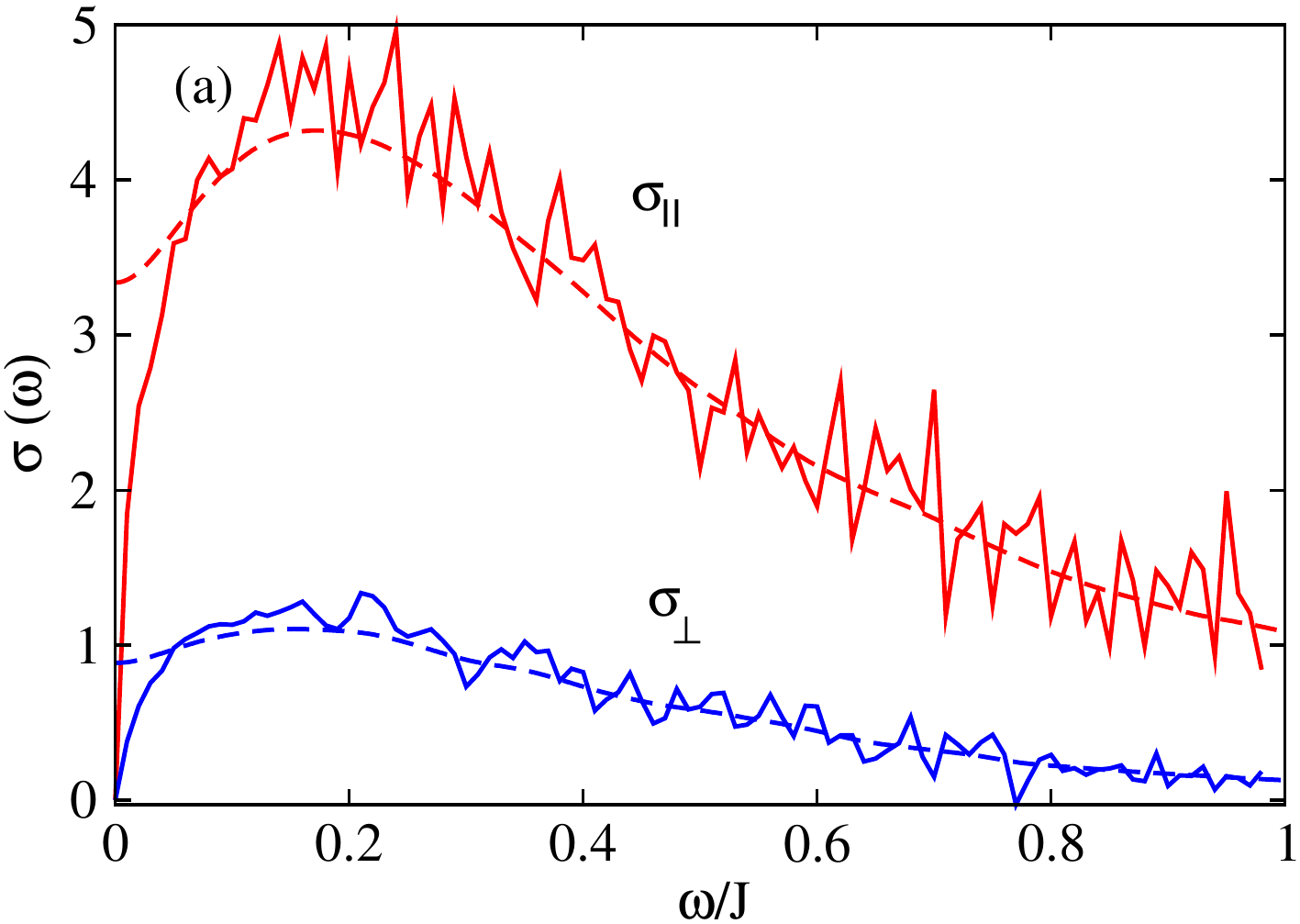}
\includegraphics[width=0.49\linewidth]{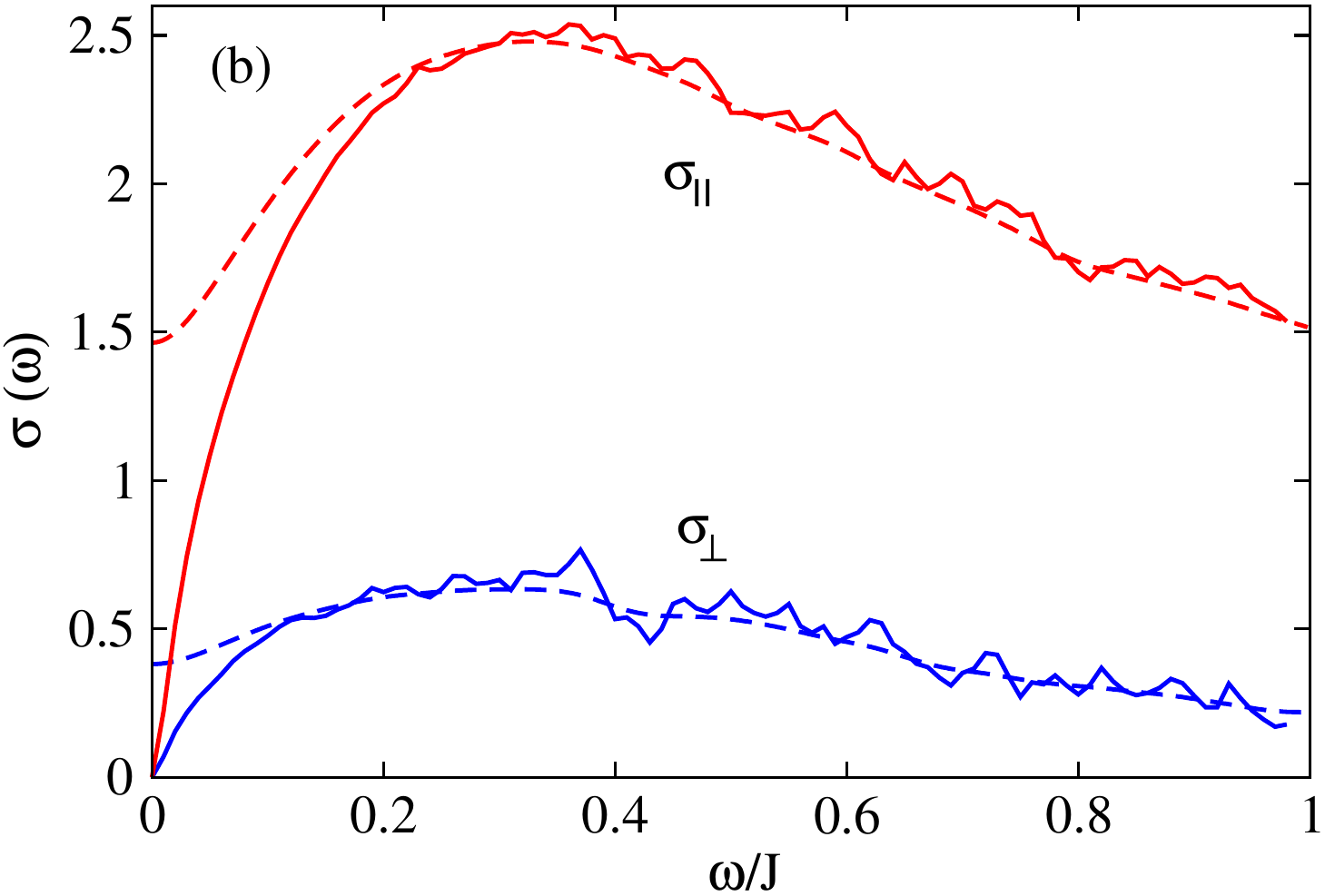}
\caption{Optical conductivity $\sigma(\omega)$ of the  2D model with uncorrelated disorder 
in both directions: parallel $\sigma_{\parallel}(\omega)$ and perpendicular $\sigma_{\perp}(\omega)$. 
(a) without extrinsic disorder ($\Delta=0$), 
(b) with extrinsic disorder ($\Delta = 0.3 J$). 
(Full line) static disorder only, (dashed line) static disorder and dynamical disorder with  $\hbar/\tau_{in}=0.05 J$.
$\sigma$ is given in unit of $\sigma_0=ne^{2}a^{2}/\hbar$, $J=0.142$\,eV, $J' = 0.028$\,eV and $T = 0.18J$.}
\label{fig:cond__t0_18_and_0}
\end{figure}
The optical conductivities of the $2D$ model are shown figures \ref{fig:cond_2D_non_and0} and \ref{fig:cond__t0_18_and_0}. In the limit of static disorder  and in the absence of extrinsic disorder the localisation peak appears around $\omega^{*}\simeq 4$ meV for $T=0.05 J$. Under the effect of thermal disorder (increase of temperature) this peak enlarges progressively and shifts toward higher frequencies. For  $T=0.18J$, one get $\omega^{*}\simeq 15$\,meV. In this case the maximum conductivity is $\sigma(\omega^{*})\simeq 15\sigma_0$ (figure \ref{fig:cond_2D_non_and0}). 

The anisotropy of the uncorrelated hopping model between the two directions of transport $\vec{a}$ and $\vec{b}$ is shown in figure  \ref{fig:cond__t0_18_and_0} by comparing the parallel conductivity, $\sigma_{\parallel}$, and the perpendicular conductivity, $\sigma_{\perp}$). The anisotropy of the conductivity ${\sigma_{\parallel}}/{\sigma_{\perp}}\simeq 3$, which is in good agreement with the experimental results which give also a value close to $3$ \cite{6exper2}. 
{
Moreover we have checked that this ratio is almost independent of the intensities of extrinsic disorder $\Delta$ and  temperature $T$.
}

\subsection{Comparison between $1D$ and $2D$ models}

\begin{figure}[h!]
\centering
\includegraphics[width=0.5\linewidth]{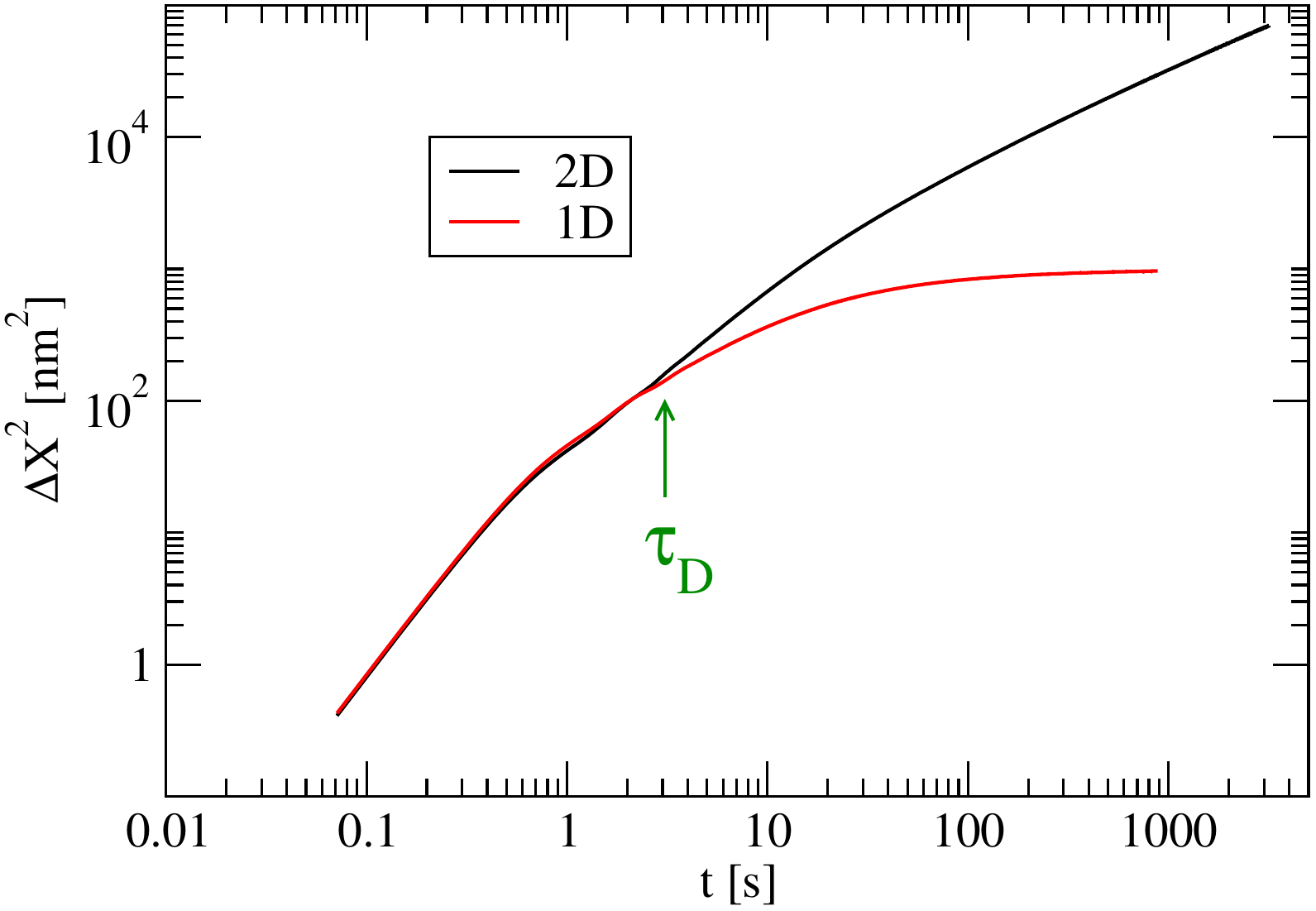}
\caption{Average square spreading $\Delta X^{2}$ vesus time $t$ for 1D model ($J' = 0$) and 2D model ($J' = 0.028$\,eV), at temperature $T=0.2 J$, with extrinsic disorder $\Delta = 0.3 J$ and $J=0.142$\,eV.}
\label{6fig:DX2_2D-1D-T02-and03}
\end{figure}

In this paragraph we compare the results of both modelisations (1D and 2D) by focusing on  the spreading $\Delta X^{2}(t)$ and the optical conductivity for a static system (i.e. no dynamical disorder). The only difference here is in the fact that for the 2D model we take into account the inter-chain coupling $J_{1}=28$\,meV which is much smaller than the intra-chain coupling $J=142$\,meV. Figure \ref{6fig:DX2_2D-1D-T02-and03} represents  the  spreading $\Delta X^{2}(t)$ for both models. One sees that for small times both propagations are ballistic and equal. Yet after a time $\tau_{D}$  the propagation in the 1D model is smaller as expected. Indeed localisation effects are stronger in 1D than in 2D.

\begin{figure}[h!]
\centering
\includegraphics[width=0.45\linewidth]{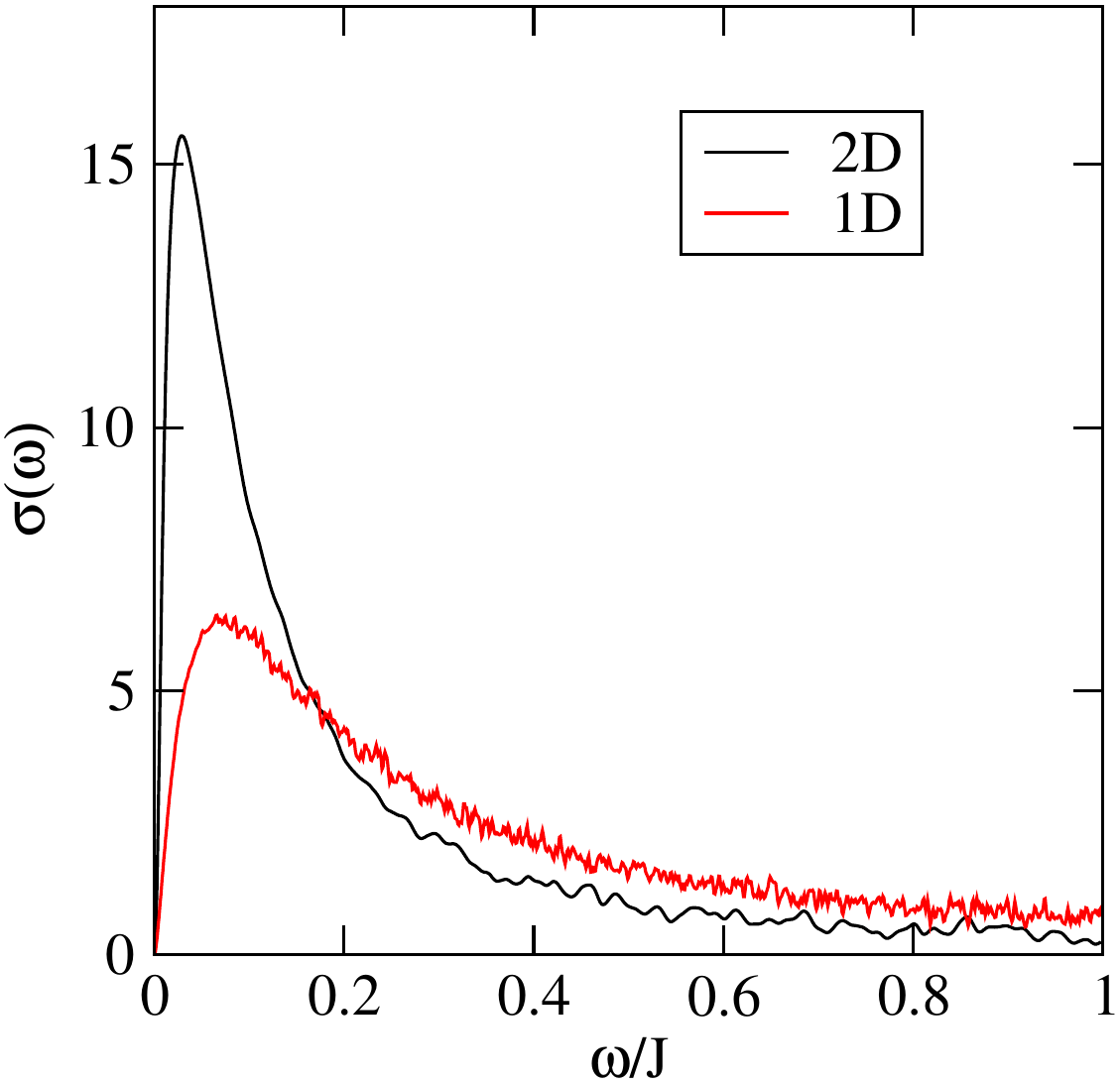}
\caption{Optical conductivity along $x$ axis, $\sigma_\parallel$, of rubrene in the 1D model ($J' = 0$) and the 2D model ($J' = 0.028$\,eV), at temperature $T=0.12 J$, with only intrinsic disorder. $\sigma$ is versus $\sigma_0=ne^{2}a^{2}/\hbar$ and $J=0.142$\,eV.}

\label{fig:Cond_2d1d}
\end{figure}

Figure \ref{fig:Cond_2d1d}  shows the difference of the optical conductivities in 1D and 2D. At large values of the frequency $\omega$ the two conductivities are comparable. This is consistent with the fact that the short time regimes for $\Delta X^{2}(t)$ are nearly identical. For smaller frequencies (i.e. larger time scales) the $\sigma_{2D}(\omega)\gg\sigma_{1D}(\omega)$ which is consistent with the fact that the diffusion satisfies $\Delta_{2D} X^{2}(t)\gg\Delta_{1D} X^{2}(t)$ for $t>\tau_{D}$. In particular the maximum value of $\sigma_{2D}(\omega)$ is about three times the maximum value of $\sigma_{1D}(\omega)$. At zero frequency both conductivities tend to zero because of Anderson localisation for static systems. Note that if we introduce a dynamical disorder characterised by an inelastic scattering time 
we expect that the diffusivity (and therefore the mobility) will be higher in the 2D model than in the 1D model. Indeed the diffusivity is given by:
\begin{equation}\label{mobility2}
D= \frac{L^{2}(\tau_{in})}{2 \tau_{in}} \, ,
\end{equation}  
where $\tau_{in}$ is the inelastic scattering time (which is of the order of the period of the inter-molecular vibrations) and $L^{2}(\tau_{in})$ is an average of the spreading $\Delta X^{2}(t)$ in the static structure over a characteristic time $\tau_{in}$. Therefore as soon as $\tau_{in}>\tau_{D}$ (which is expected) the diffusivity and the mobility are greater in the 2D model than in the 1D model.

To conclude the diffusion along the chains in the 2D model is much greater than in the 1D model except at short times $t<\tau_{D}$. This may appear surprising in view of the relatively small values of the inter-chain hopping terms (about five times smaller than the intra-chain term). This implies that the 1D model underestimates the conductivity and the mobility along the chains when compared to the more precise 2D model. This conclusion is also consistent with that given in Ref. \cite{Fratini17} where it is shown that the mobilities (in any directions)  are optimized in systems where the intra-chain and inter-chain transfer integrals are comparable.

\section{Conclusion}
{
Developing theories and understanding of  the electronic transport properties of organic semi-conductors is still a challenge. In the case of hole doped rubrene the best theory so far seems to be the transient localization scenario \cite{6Ciuchi2011,6Fartini2016}. In this scenario there is not the formation of a polaronic state, and the transient localisation scenario  emphasises the role of disorder that is due either to thermal motion or to impurities and chemical disorder. This also is in accordance with the fact that the mobility is relatively high for p-doped rubrene and much higher than for the n-doped case.
}

{
We have studied the electronic transport for hole doped rubrene within a tight-binding model that retains only one HOMO orbital per molecule. These HOMO orbitals are electronically coupled due to the proximity of the molecules and the hopping matrix elements are time dependent, with strong relative variations, due to the motion of molecules. The most important coupling is along the 1D chains of rubrene molecules but we have also considered the effect of the inter-chain coupling and we have analysed in detail its consequences  on the transport properties. To this aim we have considered different models that neglect (1D models) or fully include (2D models) this interchain coupling.  The 2D models are able to reproduce the main features such as localisation peak of the optical conductivity, and value of the zero frequency mobility.  The mobility anisotropy which is of the order of 2-4 between the direction along the chains and perpendicular to the chains, is in good agreement with the experimental value \cite{6exper2}. The comparison of the 2D and 1D models shows that the 1D model underestimate the mobility and the ac-conductivity along the chains. Our study shows the importance of the coupling between chains of rubrene molecules despite the relatively low value of the inter-chains coupling. Therefore we conclude that theoretical  treatment of transport in rubrene that will be developed in the future must take into account the interchain coupling.  Our results are  also consistent with those in Ref. \cite{Fratini17} where it is shown that the mobilities (in any directions)  are optimized with systems where the intra-chain and inter-chain transfer integrals are comparable.}

{
The numerical approach  used here to compute the optical conductivity can be applied for other organic semi-conductor with hight mobility, such as  naphtalene and anthracene \cite{6Karl2003,6Karl2001}, i.e. when the transient localisation scenario is available.
}

\section*{Acknowledgements}

The authors wish to thank Simone Fratini, Sergio Ciuchi and Pascal Qu\'emerais
for fruitful discussions.
The numerical calculations have been performed 
at  Institut N\'eel, Grenoble,
and at the Centre de Calculs (CDC),
Universit\'e de Cergy-Pontoise.
This work was supported by the Tunisian French Cooperation Project (Grant No. CMCU 15G1306),
the Institut des \'Etudes Avanc\'ees (IEA - Universit\'e de Cergy-Pontoise)
and the ANR project J2D (ANR-15-CE24-0017).




\externalbibliography{yes}
\bibliography{rubrene_biblio}




\end{document}